\documentclass[reprint,prb,superscriptaddress,showpacs]{revtex4-2}%
\usepackage{graphicx,epstopdf}
\usepackage{color}
\usepackage{amsmath}
\usepackage{amsfonts}
\usepackage{amssymb}
\usepackage{framed}
\usepackage{float}
\usepackage{url}
\usepackage[normalem]{ulem}
\providecommand{\U}[1]{\protect\rule{.1in}{.1in}}
\makeatletter
\renewcommand*{\fnum@figure}{{\normalfont\bfseries \figurename~\thefigure}}
\renewcommand*{\@caption@fignum@sep}{\textbf{ : }}
\makeatother
\usepackage{hyperref}
\hypersetup{
    colorlinks=true,
    linkcolor=blue,
    filecolor=magenta,      
    urlcolor=cyan,
}
\newcommand{\SM}[1]{\textcolor{red}{#1}}
\begin{document}

\title{Domain wall induced topological Hall effect in the chiral-lattice ferromagnet Fe$_x$TaS$_2$}

\author{Sk Jamaluddin}
\email{Corresponding author: sjamalud@nd.edu}
\affiliation{Department of Physics and Astronomy, University of Notre Dame, Notre Dame, IN 46556, USA}
\affiliation{Stavropoulos Center for Complex Quantum Matter, University of Notre Dame, Notre Dame, IN 46556, USA}
\author{Warit Nisaiyok}
\affiliation{Department of Physics and Astronomy, University of California, Riverside, CA 92521, USA}
\author{Yu Zhang}
\affiliation{Materials Science Division, Argonne National Laboratory, Lemont, IL, 60439, USA}

\author{Hari Bhandari}
\affiliation{Department of Physics and Astronomy, University of Notre Dame, Notre Dame, IN 46556, USA}
\affiliation{Stavropoulos Center for Complex Quantum Matter, University of Notre Dame, Notre Dame, IN 46556, USA}

\author{Brian A. Francisco}
\affiliation{Department of Physics and Astronomy, University of California, Riverside, CA 92521, USA}
\author{Peter E. Siegfried}
\affiliation{Department of Physics and Astronomy, George Mason University, Fairfax, VA 22030, USA}

\author{{Fehmi Sami Yasin}}
\affiliation{Center for Nanophase Materials Sciences, Oak Ridge National Laboratory, Oak Ridge, TN 37830, USA}

\author{{Tianyi Wang}}
\affiliation{Department of Physics, University of Michigan, Ann Arbor, MI 48109, USA}

\author{Abhijeet Nayak}
\affiliation{Department of Physics and Astronomy, University of Notre Dame, Notre Dame, IN 46556, USA}
\affiliation{Stavropoulos Center for Complex Quantum Matter, University of Notre Dame, Notre Dame, IN 46556, USA}
\author{{Mohamed El Gazzah}}
\affiliation{Department of Physics and Astronomy, University of Notre Dame, Notre Dame, IN 46556, USA}
\affiliation{Stavropoulos Center for Complex Quantum Matter, University of Notre Dame, Notre Dame, IN 46556, USA}
\author{Resham Babu Regmi}
\affiliation{Department of Physics and Astronomy, University of Notre Dame, Notre Dame, IN 46556, USA}
\affiliation{Stavropoulos Center for Complex Quantum Matter, University of Notre Dame, Notre Dame, IN 46556, USA}
\author{{June Ho Yeo}}
\affiliation{Department of Physics, University of Michigan, Ann Arbor, MI 48109, USA}
\author{{Liuyan Zhao}}
\affiliation{Department of Physics, University of Michigan, Ann Arbor, MI 48109, USA}

\author{J. F. Mitchell}
\affiliation{Materials Science Division, Argonne National Laboratory, Lemont, IL, 60439, USA}
\author{Yong-Tao Cui}
\affiliation{Department of Physics and Astronomy, University of California, Riverside, CA 92521, USA}
\author{Nirmal J. Ghimire}
\email{Corresponding author: nghimire@nd.edu}
\thanks{\\Notice: This manuscript has been authored by UT-Battelle, LLC, under contract DE-AC05-00OR22725 with the US Department of Energy (DOE). The US government retains and the publisher, by accepting the article for publication, acknowledges that the US government retains a nonexclusive, paid-up, irrevocable, worldwide license to publish or reproduce the published form of this manuscript, or allow others to do so, for US government purposes. DOE will provide public access to these results of federally sponsored research in accordance with the DOE Public Access Plan (https://www.energy.gov/doe-public-access-plan).}
\affiliation{Department of Physics and Astronomy, University of Notre Dame, Notre Dame, IN 46556, USA}
\affiliation{Stavropoulos Center for Complex Quantum Matter, University of Notre Dame, Notre Dame, IN 46556, USA}

\date{\today}
\begin{abstract}
Magnetic topology and its associated emergent phenomena are central to realizing intriguing quantum states and spintronics functionalities. Designing spin textures to achieve strong and distinct electrical responses remains a significant challenge. Layered transition metal dichalcogenides offer a versatile platform for tailoring structural and magnetic properties, enabling access to a wide spectrum of topological magnetic states. 
Here, we report a domain-wall–driven, large, and tunable topological Hall effect (THE) in a non-centrosymmetric intercalated transition metal dichalcogenides series Fe$_x$TaS$_2$. By systematically varying the Fe intercalation level, we exert precise control over the magnetic ground states, allowing manipulation of the topological Hall effect.
Real-space magnetic force microscopy (MFM) provides direct evidence of periodic magnetic stripe domain formation, confirming the microscopic origin of the observed topological transport phenomena. Our findings establish a promising way for tuning the topology of domains to generate substantial electromagnetic responses in layered magnetic materials.

\end{abstract}
\maketitle

\section{Introduction}\label{sec:1}

Transition metal dichalcogenides (TMDs) form a versatile class of materials that exhibit a wide range of exotic phenomena, including superconductivity \cite{revolinsky1963layer,xi2016ising}, charge density waves (CDWs) \cite{ritschel2015orbital,nakata2021robust}, Mott insulating behavior \cite{sipos2008mott,lian2024valley}, and nontrivial topological states \cite{saha2022observation}. Their layered nature, coupled with weak van der Waals bonds, allows intercalating various elements in between the layers, opening up a large and tunable materials platform. 

It is well established that intercalating 3$d$ transition metals between the layers generates a particular structure-type of the TMDs,  2H-$TX_2$ ($T$ = Nb, Ta; $X$ = S, Se), of which there are two distinct families of ordered intercalated transition metal dichalcogenides (ITMDs) \cite{van1971magnetic,parkin19803}. When the intercalate occupies one-quarter of the octahedral sites, the resulting structure is centrosymmetric, forming a $2 \times 2$ superlattice derived from the $TX_2$ unit cell and crystallizing in the hexagonal space group $P6_3/mmc$. 
Alternatively, in the case of one-third occupancy leads to a non-centrosymmetric, $\sqrt{3} \times \sqrt{3}$ superstructure, crystallizing in the chiral space group $P6_322$. 

\begin{figure*}[!ht]
\begin{center}
\includegraphics[width=0.7\linewidth]{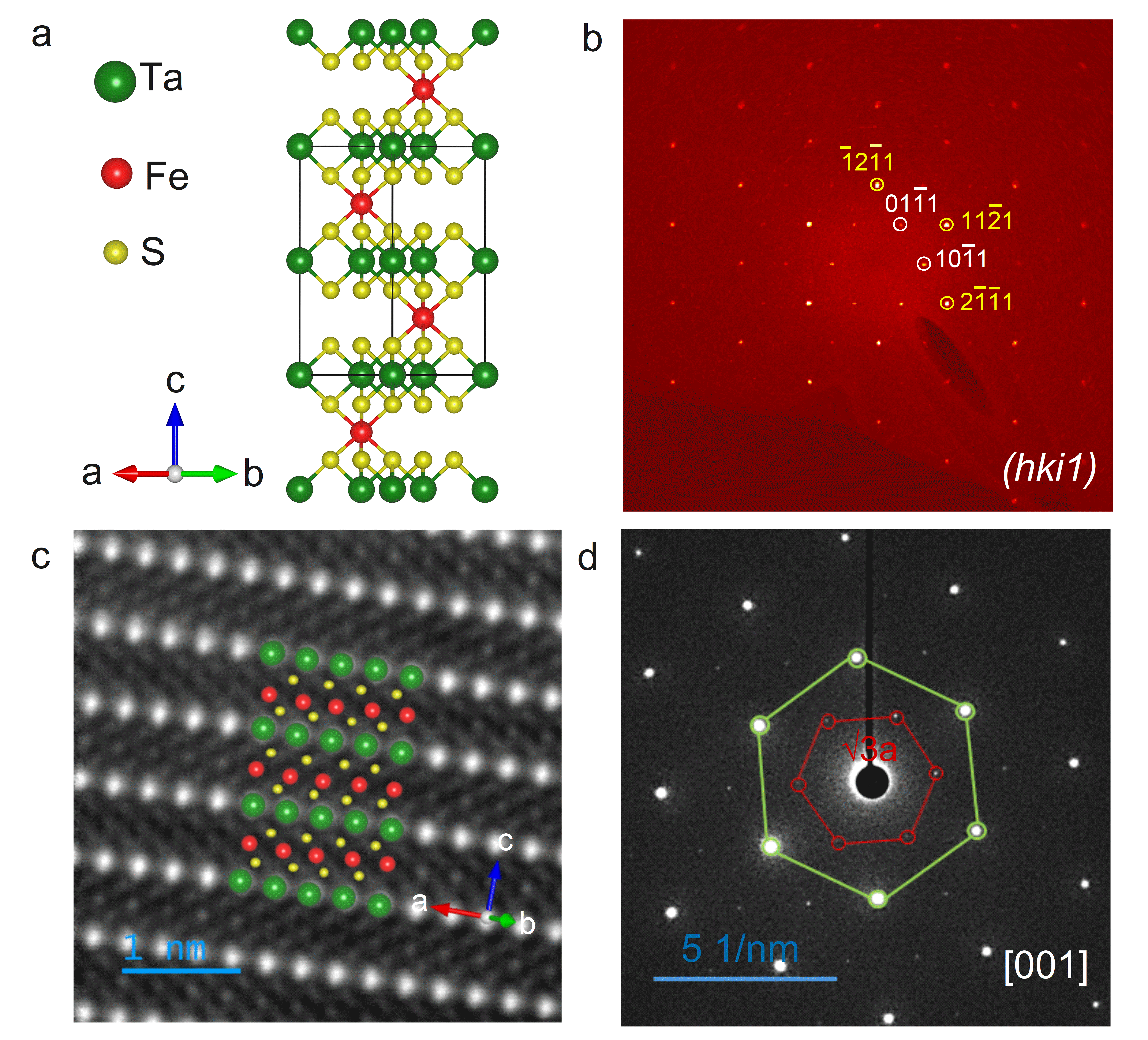}
    \caption{\small a, Schematic of the crystal structure of Fe$_{1/3}$TaS$_2$. Green, red and yellow balls represent tantalum (Ta), iron (Fe) and sulfur (S) atoms, respectively. b, Precession image in the $(hki1)$ plane for $x \approx$ 0.30 recorded by single crystal X-ray diffraction. Strong reflections (yellow circles) are associated with the underlying 2H-TaS$_2$ structure, while weak reflections (white circles) correspond to the $\sqrt{3} \times \sqrt{3}$ ordered superlattice.
    c, Atomic resolution high-angle annular dark field scanning transmission electron microscopy (HAADF-STEM) image for  $x \approx$ 0.30 along the [01$\bar{1}$0] zone axis. The inset shows the atomic arrangements. d, Selected area electron diffraction (SAED) pattern along the [001] for $x \approx$ 0.30. $\sqrt{3} \times \sqrt{3}$ superstructure unit cell is highlighted with a red hexagon, while the unit cell corresponding to TaS$_2$ is outlined by the green hexagon.}
    \label{Fig1}
    \end{center}
\end{figure*}

The physical properties of these ITMDs are significantly influenced by the nature of the intercalated transition metal. For example, Cu intercalation in TiSe$_2$ induces superconductivity \cite{morosan2006superconductivity,barath2008quantum}. Cr$_{1/3}$NbS$_2$ hosts a chiral helimagnetic structure \cite{togawa2012chiral,ghimire2013magnetic}, while Ni$_{1/3}$NbS$_2$ forms a helical antiferromagnetic state \cite{an2023bulk}. Likewise, Co$_{1/3}$NbS$_2$ \cite{ghimire2018large,tenasini2020giant,kirstein2025topological} and Co$_{1/3}$TaS$_2$ \cite{takagi2023spontaneous,park2023tetrahedral,kruppe2025anisotropic} show a large anomalous Hall effect, associated with the non-coplanar magnetic textures. The collinear antiferromagnetism in Fe$_{1/3}$NbS$_2$ is switchable by electric current \cite{nair2020electrical,haley2023long}. Furthermore, recent findings of altermagnetism in Co$_{1/4}$NbSe$_2$ and Co$_{1/4}$TaSe$_2$ further enrich the diversity of these intercalated materials \cite{regmi2025altermagnetism,dale2024non,sakhya2025electronic,sprague2025observation,mandujano2024itinerant}. 

Notably, Fe intercalation into 2H-TaS$_2$ (Fe$_x$TaS$_2$) exhibits distinct behavior among intercalated transition metal dichalcogenides (ITMDs). Both intercalated phases—the centrosymmetric 1/4 phase and the noncentrosymmetric 1/3 phase—display easy-axis ferromagnetism along the $c$-axis \cite{narita1994preparation, morosan2007sharp, mangelsen2020large, zhang2019critical}. The noncentrosymmetric structure of Fe$_x$TaS$_2$ (for $x > 0.26$) allows the emergence of a Dzyaloshinskii–Moriya interaction (DMI), whereas the centrosymmetric structure (for $x \leq 0.26$) is not expected to host such an effect. Furthermore, crystallographic defects in off-stoichiometric compositions, those lying between the ideal 1/4 and 1/3 Fe contents, may give rise to interesting transport phenomena \cite{chen2016correlations}. The compositional tunability, coupled with potential disorder-mediated effects and the continuous evolution from a noncentrosymmetric to a centrosymmetric structure while maintaining easy-axis magnetic anisotropy, makes the Fe$_x$TaS$_2$ series a unique platform for exploring intriguing magnetic and magnetotransport properties.

Here, we report the emergence of a topological Hall effect (THE) in the noncentrosymmetric structure, associated with the formation of particular striped domain walls. In contrast, the centrosymmetric composition lacks both the striped domains and the corresponding THE, providing compelling evidence for a domain-wall-induced topological Hall effect. The THE arises when charge carriers traverse a magnetic texture with finite scalar spin chirality, acquiring a Berry curvature that acts as an effective magnetic field \cite{neubauer2009topological}. As a result, the carriers exert a spin-transfer torque on the magnetic structure. Magnetic skyrmions, which can be driven by ultra-low current densities even in bulk systems, have therefore attracted significant attention for potential spintronic applications \cite{jonietz2010spin,schulz2012emergent}. Ferromagnetic domain walls are already used in spintronic devices, but they typically require relatively high current densities for manipulation \cite{parkin2008magnetic,luo2020current}. The domain walls observed in Fe$_x$TaS$_2$, capable of producing a topological Hall response, may thus provide insight into emergent physics relevant to low-power spintronic devices. 

\begin{figure*}[!ht]
\begin{center}
	\centering
	\includegraphics[angle=0,width=16cm,clip]{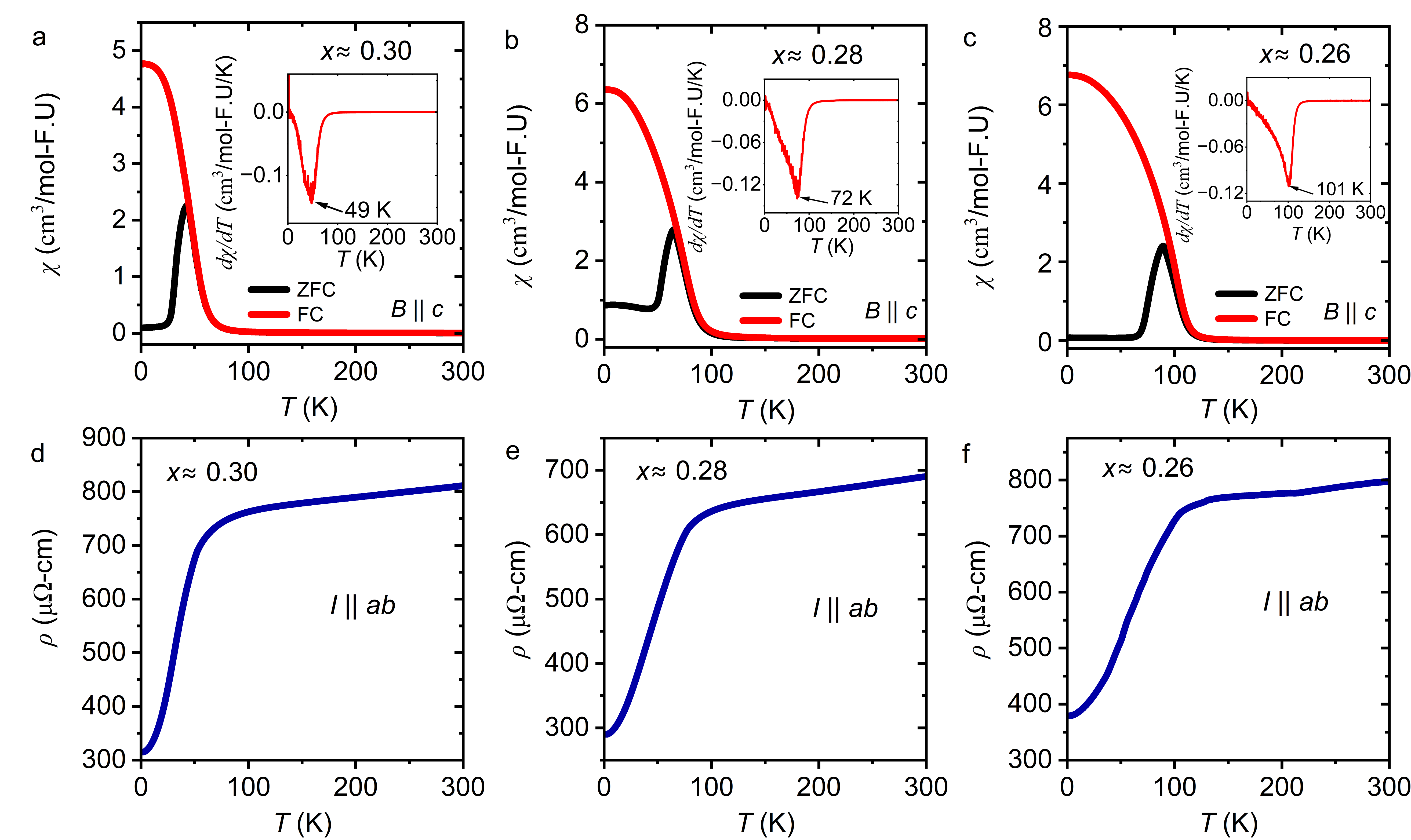}
	\caption{ a-c, Temperature-dependence of magnetic susceptibility [$\chi(T)$] measured with applied field of 0.1 T along the $c$-axis for Fe$_{x}$TaS$_2$ ( $x \approx$ 0.30, 0.28, 0.26). Insets show the temperature derivative of the magnetic susceptibility. d-f, Zero-field longitudinal resistivity $[\rho(T)]$ as a function of temperature with current applied along $ab$ ($I||ab$) for Fe$_x$TaS$_2$ ($x \approx$ 0.30, 0.28, 0.26).
		\label{Fig2}}
    \end{center}
\end{figure*}

 \section{Results and Discussion}\label{sec:2}

 Taking advantage of the compositional tunability of Fe$_x$TaS$_2$, we synthesized single crystals with three different Fe concentrations: $x \approx 0.26$, 0.28, and 0.30 (see Supplementary Fig.\SM{S1} for compositional analysis). Because of the complexity associated with the layered nature of the crystals, we use multimodal approach for structural characterization, including single-crystal X-ray diffraction (SC-XRD) and high-resolution scanning transmission electron microscopy (HAADF-STEM) (see Methods for details on structural characterization). For the $x \approx 0.30$ composition, high-quality SC-XRD data enabled a successful refinement of the crystal structure in the noncentrosymmetric $P6_3$22 space group, a schematic of which is shown in Fig.\hyperref[Fig1]{1a}. A diffraction pattern from the $(hki1)$ scattering plane is shown in Fig.\hyperref[Fig1]{1b}, with additional refinement details and other scattering planes provided in the Supplementary section \SM{S2}. The strong reflections marked by yellow circles correspond to the underlying 2H-TaS$_2$ structure, whereas the weak reflections highlighted by white circles indicate the presence of superlattice spots associated with the $\sqrt{3} \times \sqrt{3}$ superstructure. In contrast,  SC-XRD data for the $x \approx 0.28$ and $x \approx 0.26$ samples revealed a more complex structural behavior. In particular, for $x \approx 0.26$, both specimens in which the intercalated Fe was ordered into a $2 \times 2$ superlattice and disordered (characterized by a $1 \times 1$ unit cell) were identified by single crystal XRD (see \SM{S2} for details).
 
To overcome this limitation, we used STEM and TEM electron diffraction techniques to characterize the crystal structure of all three compositions. A HAADF-STEM image of $x \approx 0.30$ along the [01$\bar{1}$0] zone axis is shown in Fig.\hyperref[Fig1]{1c}. The Fe atoms (represented by red spheres) are uniformly intercalated between the TaS$_2$ layers, with slightly reduced intensity at certain positions, likely due to Fe deficiency, and no observable defects are present in the probed region. The atomic arrangement is consistent with the noncentrosymmetric $P6_3$22 structure illustrated by the crystal structure overlayed in Fig.\hyperref[Fig1]{1c}. Figure \hyperref[Fig1]{1d} shows SAED pattern for $x \approx 0.30$ along the [001]. The diffraction pattern clearly reveals the formation of $\sqrt{3} \times \sqrt{3}$ superstructure. The strong diffraction spots (highlighted by green circles) correspond to the underlying 2H-TaS$_2$ lattice, while the weaker reflections (highlighted by red circles) are associated with the $\sqrt{3} \times \sqrt{3}$ superstructure. The SAED patterns for $x\approx 0.28$ and $x\approx 0.26$ are shown in Supplementary Fig. \SM{S3}. For $x\approx 0.28$, the Fe intercalation forms $\sqrt{3} \times \sqrt{3}$ superstructures corresponding to the P6$_3$22 space group, while in $x\approx 0.26$ it forms $2 \times 2$ superstructures which belong to the centrosymmetric P6$_3$/mmc space group. It is, however, to be noted here that, these off stoichiometric samples have inherent structural disorder  unlike the stoichiometric end compounds.

\begin{figure*}[tb!]
	\centering
	\includegraphics[angle=0,width=16cm,clip]{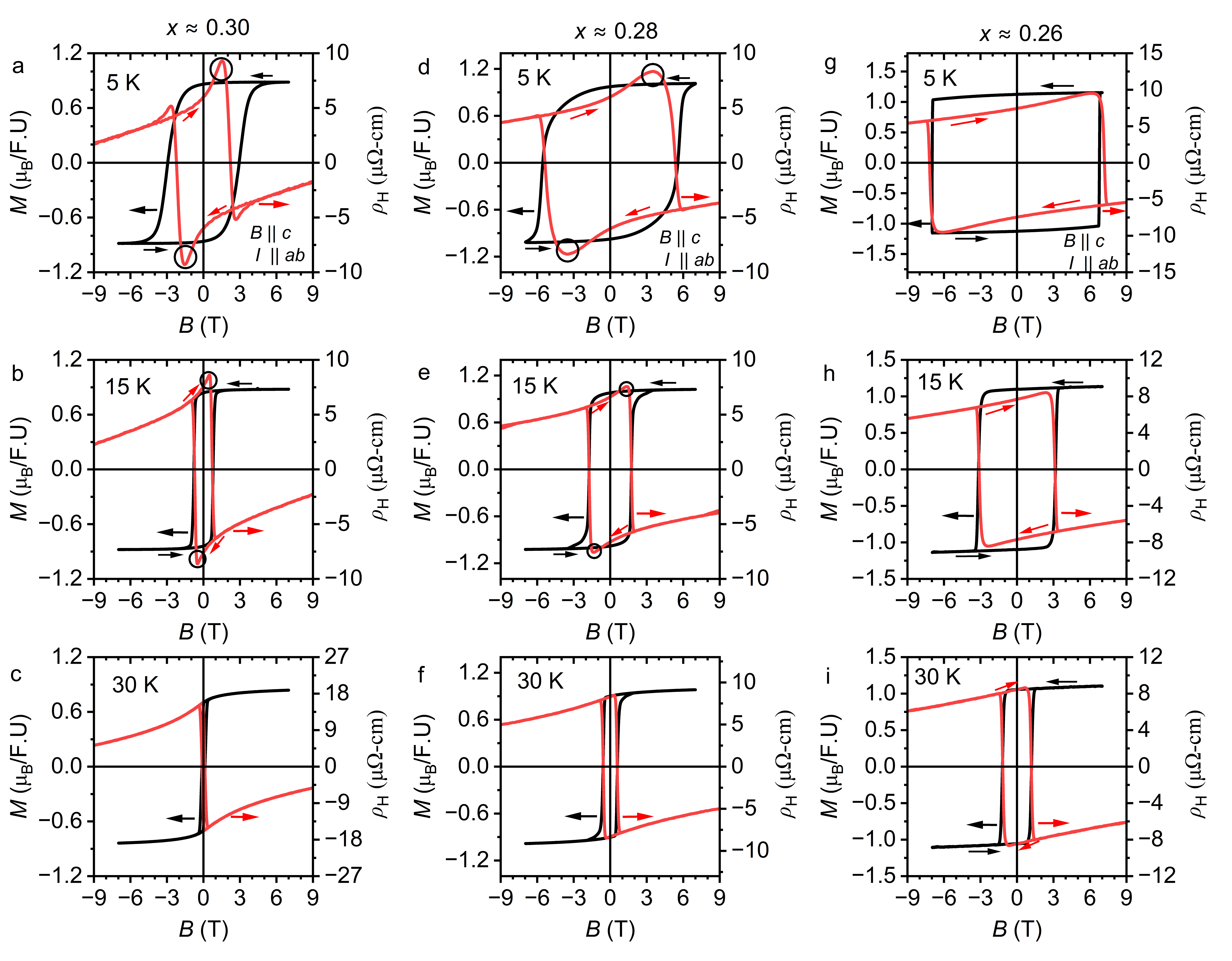}
	\caption{ a-i, Field-dependent magnetization [$M(B)$] (black curve, plotted on the left axis) and Hall resistivity [$\rho_{H}(B)$] (red curve, plotted on the right axis) measured at selected temperatures for Fe$_{x}$TaS$_2$ with  $x \approx 0.30$  (a-c), $x \approx 0.28$ (d-f), $x \approx 0.26$ (g-i). The applied field is along the c-axis ($B||c$) and the current is along the $ab$ plane ($I||ab$). The dip-like topological features are highlighted by black circles. The directions of the magnetic field sweeps are indicated by arrows.
		\label{Hall}}
\end{figure*}

Additionally, we examined the state of inversion symmetry in $x \approx 0.30$ sample optically using rotation anisotropy second harmonic generation (RA-SHG) \cite{jin2020observation,ahn2024electric} measurements as shown in Supplementary Fig. \SM{S4}. The enhancement of SHG signal, together with the preservation of the three-fold rotational symmetry ($C_3$), agrees with the breaking of spatial inversion symmetry in this composition.

We also characterized the samples for their magnetic properties.  The temperature-dependent DC magnetic susceptibility for Fe$_{x}$TaS$_2$ with $x \approx 0.30, 0.28, 0.26$ exhibits ferromagnetic behavior as shown in Figs. \hyperref[Fig2]{2a-c}. $T_{\text{C}}$ for  $x \approx 0.30$ is 49 K and increases to 72 K for $x \approx 0.28$ and for $x \approx 
0.26$ it further increases to 101 K, consistent with a solid solution of the intercalate Fe between the 1/3 and 1/4 endpoints \cite{eibschutz1981ferromagnetism,narita1994preparation,mangelsen2020large,chen2016correlations,miao2022anisotropic}. The temperature derivative of the magnetic susceptibility is presented in the inset of each corresponding susceptibility plot.  These show a clear transition near $T_{\text{C}}$ for all compositions without any feature at any other temperature, which argues against the presence of any secondary phases or significant compositional variation within a single crystal, which would have resulted in multiple transitions in the susceptibility data.

\begin{figure*}[tb!]
	\centering
	\includegraphics[angle=0,width=17cm,clip]{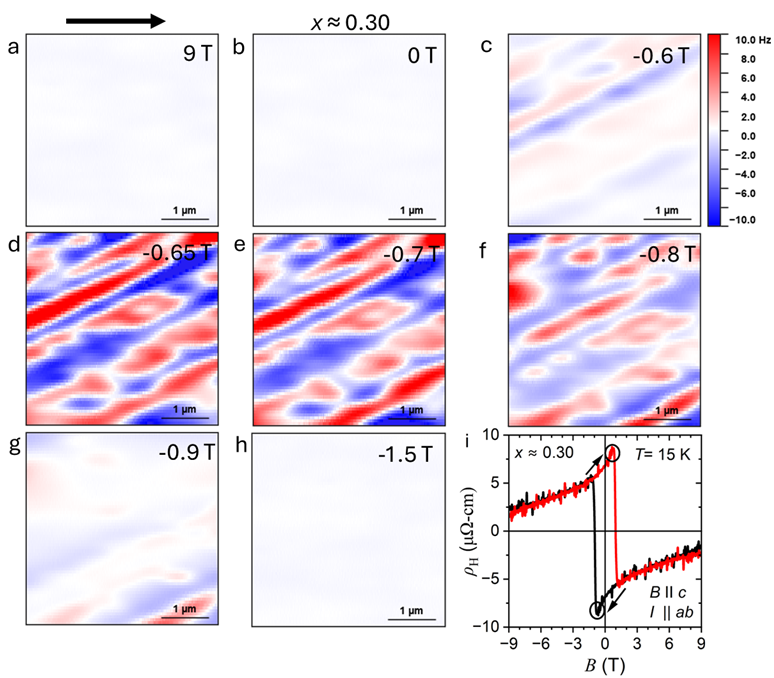}
	\caption{ a-h, Field evolution of magnetic domains for $x \approx 0.30$ at 15 K.  i, In-situ Hall resistivity measured simultaneously for $x \approx 0.30$ at 15 K. The THE features are outlined by the black circles in the Hall resistivity. The color bar represents the change in resonance frequency of the cantilever’s oscillation due to magnetic interactions between the MFM tip and the sample. The black arrows indicate the direction of the magnetic field sweeps.
		\label{MFM Fe 0.30}}
\end{figure*}

Similarly, the resistivity exhibits a sharp drop below $T_C$ for each composition, as shown in Figs. \hyperref[Fig2]{2d-f}, further supporting the presence of a single magnetic transition. This resistivity behavior and also the magnetoresistance behaviors shown in Fig. \SM{S11} are consistent with that reported for similar compositions in the literature \cite{chen2016correlations,hardy2015very}. For consistency, we measured all the magnetic, magnetotransport and single crystal diffraction on the same piece (see Methods section for details). 
 
We now focus on the magnetic field ($B = \mu_0 H$) dependence of the magnetic moment (or simply magnetization) $M$, (black curve), and Hall resistivity, $\rho_H$ (red curve), for Fe$_x$TaS$_2$ samples, measured at various temperatures, as shown in Fig.\ref{Hall}. These measurements were performed with the magnetic field applied along the $c$-axis ($B \parallel c$), while for the Hall measurements, the electric current ($I$) was applied within the $ab$-plane.

We first examine the $M$ vs. $B$ curves for all three samples at 5 K, shown in Figs. \hyperref[Hall]{3a}, \hyperref[Hall]{3d}, and \hyperref[Hall]{3g}. Across all compositions, the hysteresis loop spans a similar width of approximately $\pm6$ T. However, the saturation magnetization ($M_S$) increases as the Fe concentration ($x$) decreases, and the hysteresis loops become increasingly square-shaped. With increasing temperature, the hysteresis width narrows, although $M_S$ remains nearly constant below the Curie temperature ($T_C$) for each composition, as shown in Figs. \hyperref[Hall]{3b-i}, with additional data at higher temperatures provided in Supplementary Figs. \SM{S5} - \SM{S7}.

We now turn to the Hall resistivity, presented in Fig. \ref{Hall}. For $x \approx 0.30$, at 5 K, the Hall resistivity initially shows a linear dependence on the magnetic field as it is swept from +9 T toward 0 T (Fig. \hyperref[Hall]{3a}). However, as the field is further reduced below 0 T, a dip-like anomaly (highlighted by the black circle) appears. Upon continuing the sweep into negative fields, the Hall resistivity changes sign and exhibits a hump-like feature. At higher negative fields, the Hall resistivity re-aligns with the magnetization trend.

The Hall resistivity $\rho_H$ of a ferromagnetic system can be expressed as $\rho_H = R_0B + R_sM + \rho^T_H$, where the first, second, and third terms represent the normal, anomalous, and topological Hall contributions, respectively. The normal Hall effect exhibits a linear dependence on the external magnetic field and corresponds to the linear behavior observed in the region of saturated magnetization. The anomalous Hall term follows the magnetization curve. The dip and hump in $\rho_H$ represent the topological Hall effect ($\rho^T_H$) and have been observed in $x=$ 0.28 and 0.29 samples previously \cite{cai2019anomalous,zheng2021tailoring}. The amplitude of $\rho^T$ decreases with increasing temperature, as shown in Figs. \hyperref[Hall]{3b} and \hyperref[Hall]{3c}, and vanishes at and above 30 K.

Although normal and anomalous Hall components are present in all compositions, the topological Hall contribution, $\rho^T_H$, systematically decreases with decreasing Fe content. At 5 K, this trend is evident in Fig. \hyperref[Hall]{3d} for $x \approx 0.28$, and it vanishes completely for $x \approx 0.26$ (Fig. \hyperref[Hall]{3g}). This observation supports the notion that the topological Hall effect is closely linked to the strength of DMI \cite{zheng2021tailoring}. Similar to the $x \approx 0.30$ composition, the $x \approx 0.28$ sample exhibits a temperature-dependent suppression of $\rho^T_H$, which disappears at and above 30 K, as shown in Figs. \hyperref[Hall]{3e-f} and Supplementary \textcolor{red}{Fig. S6}. In contrast, the $x \approx 0.26$ composition displays only the normal and anomalous Hall components across the entire temperature range [Figs. \hyperref[Hall]{3g-i} and Supplementary Fig. \SM{S7}]. Although $\rho^T_H$ is observed in the $x \approx 0.30$ and $x \approx 0.28$ samples, the longitudinal resistivity [$\rho(B)$] exhibits a nearly identical behavior across all compositions (see Supplementary Figs. \SM{S8-S11} ).

\begin{figure*}[tb!]
	\centering
	\includegraphics[angle=0,width=16cm,clip]{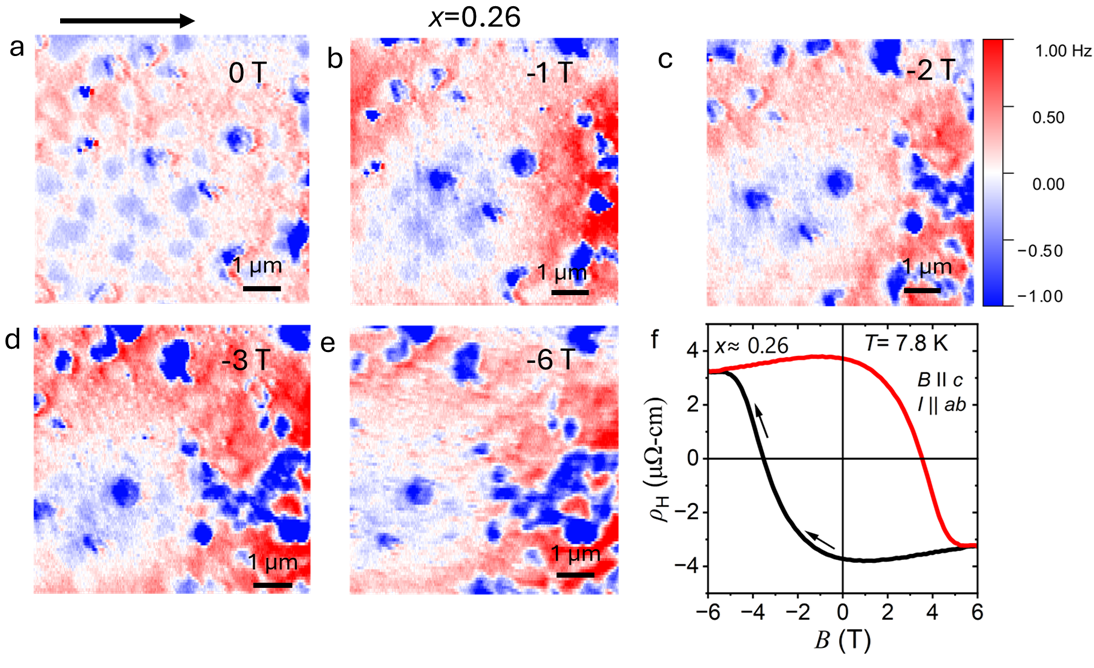}
	\caption{ a-e, Field evolution of magnetic domains at 7.8 K for $\approx 0.26$.  f, Simultaneously measured in-situ Hall resistivity at 7.8 K on the same sample. The black arrows indicate the direction of the magnetic field sweeps.
		\label{MFM Fe 0.26}}
\end{figure*}

A similar THE, previously reported in heterostructure or thin-film systems, has been attributed to the formation of skyrmion or antiskyrmion textures \cite{wu2022van,sivakumar2020topological}.
 Motivated by these findings, we performed MFM measurements on Fe$_x$TaS$_2$  samples at the two compositional extremes $x \approx 0.30$ and $x \approx 0.26$. To directly correlate the magnetic textures with transport behavior, simultaneous Hall resistivity measurements were carried out on the samples during the MFM experiments (see Methods for the details of MFM measurements). MFM detects the gradient of the magnetic stray field, which is stronger near the magnetic domain walls; hence, a large signal contrast indicates the presence of magnetic domains, and the spatial pattern reveals their morphology. Figure \ref{MFM Fe 0.30} shows a sequence of MFM images for Fe$_{0.30}$TaS$_2$ acquired at 15 K under out-of-plane ($c$-axis) magnetic fields. After saturating the magnetization at 9 T, the field was gradually swept down to -9 T, with images collected at selected field values. 

From 9 T down to slightly below zero field, a uniform magnetic contrast is observed (Figs. \hyperref[MFM Fe 0.30]{4a,b}). As the magnetic field is decreased to –0.6 T, faint stripe-like features begin to appear (Fig. \hyperref[MFM Fe 0.30]{4c}). With slightly further reduction of the field, well-defined stripe domains develop, exhibiting strong magnetic contrast (Figs. \hyperref[MFM Fe 0.30]{4d,e}). Upon decreasing the field further to -0.8 T, the stripe domains gradually shrink (Figs. \hyperref[MFM Fe 0.30]{4f,g}) and eventually disappear completely beyond –1 T, after which a uniform magnetic contrast re-emerges (Fig. \hyperref[MFM Fe 0.30]{4h} and Supplementary Fig. \SM{S12}).

Simultaneous in-situ Hall resistivity measurements (Fig. \hyperref[MFM Fe 0.30]{4i}) are consistent with the Hall data shown in Fig. \ref{Hall} (see also Supplementary Fig. \SM{S13}). Notably, the formation of stripe domains coincides precisely with the field range where the topological Hall effect is observed (highlighted by black circles in Fig. \hyperref[MFM Fe 0.30]{4i}), directly linking the topological Hall signal with the emergence of the striped domains in $x \approx 0.30$ composition. MFM images obtained during the reverse field sweep from -9 T to +9 T at 15 K are shown in Supplementary Fig. \SM{S14}. Additional MFM and in-situ Hall resistivity measurements performed at 7.8 K are presented in Supplementary Figs. \SM{S15-S16}. At this lower temperature, the stripe domains become wider while remaining confined to the same field range in which the THE appear (Supplementary Fig. \SM{S15}i).

Similar MFM measurements were performed on the $x \approx 0.26$ sample, with corresponding images obtained at 7.8 K during the negative field sweep shown in Fig. \ref{MFM Fe 0.26}. This composition, during the magnetization reversal, exhibits irregularly shaped and randomly oriented domains, showing no clear periodicity across the measured magnetic field range.
Unlike the $x \approx 0.30$ sample, no THE signal is observed here (Fig.\hyperref[MFM Fe 0.26]{5f}) in the simultaneous in-situ Hall measurement, consistent with the absence of a THE contribution in the direct Hall measurements presented in Figs. \hyperref[Hall]{3g-i}.

These results provide direct real-space evidence linking the THE to the presence of ordered striped domains in $x \approx 0.30$ composition. A plausible explanation is that the DMI is allowed in $x \approx 0.30$ but not in $x \approx 0.26$ \cite{zheng2021tailoring}. The presence of DMI lifts the degeneracy between left- and right-handed spin configurations. Consequently, the stripe domain walls observed in $x \approx 0.30$ sample are chiral, with a spin configuration that gives rise to finite scalar spin chirality \cite{schoenherr2018topological}. These chiral domain walls are therefore topological in nature and are expected to experience spin-transfer torque, enabling their motion under ultra-low current densities compared to conventional ferromagnetic domain walls.

What makes this system particularly intriguing is that, in a more ordered sample where the Fe content is closer to 1/3 \cite{mangelsen2020large}, the THE is absent. This raises the intriguing possibility that crystallographic defects, even in minute concentration, in combination with the non-centrosymmetric structure, play an essential role in stabilizing chiral domain walls. Although this aspect has not yet been widely recognized, it appears to be a consistent feature across multiple ITMDs. For instance, in Co$_x$TaS$_2$, the stoichiometric compound ($x = 1/3$) exhibits a single magnetic transition and no anomalous Hall effect (AHE) \cite{park2024composition}, whereas samples with $x < 0.325$ display two magnetic transitions, with the AHE emerging in the low-temperature phase associated with a multi-\textit{Q} antiferromagnetic structure \cite{park2023tetrahedral,kruppe2025anisotropic}.

Our observations establish Fe$_{0.30}$TaS$_2$ as a promising bulk platform for exploring and controlling topological domain walls. The coexistence of strong spin–orbit coupling and a robust topological Hall response suggests the potential for generating substantial spin-transfer torques under applied currents. These torques can enable efficient, low-power manipulation of domain walls, providing an alternative to engineered multilayer systems or those relying on more elusive skyrmion phases. Furthermore, this finding calls for a deeper investigation into the role of site disorder and highlights the possibility of designing exotic properties across a broad family of ITMDs through deliberate defect engineering, achieved either chemically or physically, for instance, via controlled radiation damage.

\section{Conclusion}

In summary, our study demonstrates that both the magnetic structure and the corresponding Hall response in Fe$_x$TaS$_2$ can be systematically tuned through Fe intercalation. Notably, the composition with $x \approx 0.30$ exhibits a pronounced THE, which correlates directly with the emergence of periodic striped domain-wall structures observed by magnetic force microscopy within the same field range. The combined insights from transport measurements and real-space imaging provide compelling evidence that the observed THE originates from domain-wall-driven mechanisms—an uncommon phenomenon in ferromagnetic systems. These findings establish Fe$_x$TaS$_2$ as a promising platform for exploring low-power domain-wall-mediated transport and highlight the potential for designing such phenomena through defect engineering within the broader ITMD family of compounds.

\section{Methods}\label{S2}
\subsection{Single Crystal Growth}
Single crystals of various Fe intercalated Fe$_x$TaS$_2$ ($x\approx$ 0.30, 0.28, 0.26) were synthesized by the chemical vapor transport (CVT) method using iodine as the transport agent. First, polycrystalline precursor powder was prepared by thoroughly mixing the high-purity iron (Alfa Aesar 99.998\%), tantalum (Alfa Aesar 99.999\%), and sulfur (Alfa Aesar 99.9995\%) in a stoichiometric ratio. The mixture was sealed in an evacuated quartz ampule and heated at 950 $^\circ$C for one week. Subsequently, 1 g of polycrystalline powder is loaded with 0.2 g iodine in an evacuated silica tube (20 mm inner diameter, 13 cm length). The tube was then loaded in a horizontal furnace with a hot end at 900 $^\circ$C and a cold end at 800 $^\circ$C for 7 days. Subsequently, the furnace was cooled to room temperature. This resulted in several well-faceted flat plate-like crystals.

\subsection{Single Crystal XRD}
The crystal structures of Fe$_x$TaS$_2$ ($ x\approx$ 0.30 and $ \approx$ 0.26) were examined by single-crystal X-ray diffraction (SCXRD). The measurements were performed on the same samples used for the magnetization and resistivity measurements. Several crystals for each compound with a typical size of $30\times 30\times 10\ \mathrm{\mu}\mathrm{m}^3$ were separated from the bulk crystals, and data  collected using a Bruker D8 X-ray diffractometer equipped with APEX2 area detector and Mo K$\alpha$ source ($\lambda = 0.71\ \text{\AA}$). Data integration, unit cell refinement, numerical absorption corrections, and precession image generation were carried out using the APEX3 software suite. Data integrations were carried out with the SAINT program, while absorption corrections were applied using  SADABS. Precession images are generated within APEX3 to visualize reciprocal lattice planes. A custom Python script was developed to remove bad pixels from the frames prior to this step. The crystal structures were solved using the SHELXS program and subsequently refined by SHELXL refinement package \cite{sheldrick2015crystal}.

\subsection{ EDS, STEM Sample Preparation and Characterization} 
The elemental composition of all samples, measured on the same crystals used to prepare magnetic and electrical resistivity samples, was determined through energy-dispersive X-ray spectroscopy (EDS) using a field-emission scanning electron microscope (FESEM, Magellan 400) equipped with an EDS detector (Bruker). Although EDS provides an approximate composition, magnetic transition, magnetization and magnetoresistance values of these compositions (see Figs. 2, 3 and \SM{S11}) are consistent to those previously reported \cite{chen2016correlations}.

Cross-sectional and planar-view TEM lamellae were prepared using a Ga$^+$ ion-based focused ion beam (FIB) following a standard lift-out procedure. The TEM samples were prepared from the same crystals used to prepare the Hall samples for the electrical resistivity measurements. To minimize Ga$^+$ ion-induced surface damage, a final thinning of the lamellae was performed using a low-energy (2 kV) Ga$^+$ ion. High-Angle Annular Dark-Field Scanning Transmission Electron Microscopy (HAADF-STEM) images and Selected Area Electron Diffraction (SAED) patterns were recorded using a double-tilt holder on a probe-corrected Spectra 300 transmission electron microscope (Thermo Fisher Scientific, USA).

\subsection{Second harmonic generation (SHG) measurements } 

Second harmonic generation (SHG) measurements were conducted using a 200 $kHz$ pulsed laser with the center wavelength being at 800 $nm$ and the pulse duration being 50$fs$. The laser is focused on the sample surface in normal incidence reflection geometry, with a spot size of 20 $\mu m$ (full width at half maximum). The polarization of the incident light is determined by a half-wave plate, with the analyzer (ultra-broadband polarizer) co-rotating at an angle difference of 0$^\circ$ (parallel channel) or 90$^\circ$ (cross channel) from the incident polarization. The reflected SHG signal is projected onto an electron-multiplying charge-coupled device (EM-CCD), from which the intensity is obtained by integrating over the pixels.

\vspace{2mm}
\textbf{Fitting of SHG}
\vspace{1mm}
The electric field from the bulk ED SHG under point group 32 is simulated to be:

\vspace{1mm}
Parallel channel:	$ E_{x=0.30}^{SS}(\phi) = \chi_{xxx} \sin[3\phi]$

\vspace{2mm}
Cross channel: 	$E_{x=0.30}^{SP}(\phi) = \chi_{xxx} \cos[3\phi]$

\subsection{Magnetic property measurements}
Magnetic property measurements were carried out using a Quantum Design Magnetic Property Measurement System-3 (MPMS-3) with a 7 T magnet. The magnetization measurements were performed on the same Hall samples that were used to measure the electrical resistivity. Before the measurements, the electrical contacts used for electrical resistivity measurements were removed, and the epoxy used to attach them was carefully cleaned off.  
This approach ensures that any differences between the two properties are not artifacts arising from demagnetization effects or sample-to-sample variation, such as the presence of impurities or slight intercalate compositional variation. 
Both the zero field-cooled (ZFC) and field-cooled (FC) measurements were conducted while warming the samples from 1.8 K to 300 K.  The magnetization measurements were conducted by sweeping the magnetic field from $+7~\mathrm{T}$ to $-7~\mathrm{T}$, followed by a reverse sweep from $-7~\mathrm{T}$ back to $+7~\mathrm{T}$.

\subsection{Electrical transport measurements}
Electrical transport properties were measured using a Quantum Design Dynacool Physical Property Measurement System (QD-PPMS). 

Single crystals of Fe$_x$TaS$_2$ were oriented along the [001] direction and shaped into a rectangular Hall bar shape for resistivity measurements. The longitudinal resistivity and Hall resistivity were measured using the conventional 4-probe method. Electrical contacts to the crystals were made by attaching 20 $\mu m$ Pt wires with Epotek H20E silver epoxy, resulting in a contact resistance below 40 $\Omega$. An excitation current of 2 mA was applied for the electrical transport measurements. 

The longitudinal resistivity ($\rho$) was determined from the measured longitudinal resistance ($R$) using the relation $\rho=RA/l$, where $A=t\cdot w$ is the cross-sectional area of the sample ($t$ is the sample thickness, $w$ is the width), and $l$ is the distance between the voltage contacts.

The Hall resistivity, $\rho_H$, was calculated using the relation $\rho_H=R_H \cdot t$, where $R_H=V_H/I$ is the measured Hall resistance and $t$ is the sample thickness. Here, $V_H$ represents the transverse voltage developed when a perpendicular magnetic field and current are applied to the Hall sample. The developed transverse voltage $V_H$, applied field $ B$, and current $I$ are orthogonal to each other.

 Contact misalignment in the longitudinal resistivity and Hall resistivity measurements were corrected using the following symmetrization and anti-symmetrization techniques, respectively. \\

$ \rho=\frac{\rho (+B)+\rho (-B)}{2}, ~~  \rho_H=\frac{\rho_H (+B)-\rho_H (-B)}{2}$

\vspace{2mm}

 The longitudinal and Hall resistivity measurements were performed using a four-loop magnetic field sweep protocol, where the magnetic field ($B$) was first swept from $+B_{\text{max}}$ ($+9~\mathrm{T}$) to $-B_{\text{max}}$ ($-9~\mathrm{T}$), and then back from $-B_{\text{max}}$ ($-9~\mathrm{T}$) to $+B_{\text{max}}$ ($+9~\mathrm{T}$). Symmetrization and antisymmetrization were applied to data acquired during magnetic field sweeps from \( +B_{\mathrm{max}} \) to 0 T and from \( -B_{\mathrm{max}} \) to 0 T, as well as from 0 T to \( +B_{\mathrm{max}} \) and from 0 T to \( -B_{\mathrm{max}} \).
 
Magnetization and magnetotransport measurements were reproduced on multiple crystals from different growth batches to ensure consistency and reproducibility (see Supplementary Figs. \SM{S17-S19}).

\subsection{Magnetic force microscopy measurements}
 The MFM measurements were performed in a home-built low-temperature scanning probe microscope using commercial MFM probes (Bruker MFMV) with a spring constant of 3 $\text{N}\cdot \text{m}^{-1}$, a resonance frequency at $\sim 75$ kHz, and a Co–Cr magnetic coating. MFM images were taken in a constant height mode with the tip scanning plane at $\sim 100$ nm above the sample surface. The MFM signal, the change in the resonance frequency, was measured by a Nanonis SPM controller using a phase-lock loop. The magnetic moment of the probe was nominally in the order of $10^{-16} \text{A}\cdot \text{m}^2$. Samples were mounted on a holder with electrical contacts to perform transport measurement in-situ using standard lock-in techniques.

\begin{acknowledgments}
This work was primarily supported by the US Department of Energy, Office of Science, Basic Energy Sciences, Materials Science and Engineering Division. W.N., B.F. and Y.T.C. acknowledge support from the National Science Foundation grant NO. DMR-2145735. L.Z. acknowledges support from the U.S. Department of Energy (DOE), Office of Science, Basic Energy Science (BES), under award No. DE-SC0024145 (for the measurement and analysis of RA SHG). Work conducted as part of a user project at the Center for Nanophase Materials Sciences, a U.S. Department of Energy Office of Science User Facility at Oak Ridge National Laboratory.
\end{acknowledgments}

\section*{Data availability}
The authors declare that the main data supporting the findings of this study are available within the article. Additional data are available from the corresponding author upon request.

\section*{Author contributions}
N.J.G. and S.J. conceived the idea. S.J. and H.B synthesized the crystal. R.R., A. N. and M.E.G. helped with sample characterization. S.J, P.S. and H.B. measured the magnetic, and magnetotransport properties. S.J. and F.S.Y. conducted independent STEM measurements. T.W., J.H.Y. and L.Z. carried out the SHG measurements. W.N. and B. F. performed the MFM measurements under the supervision of Y.-T.C.. Y.Z. and J.F.M. contributed single-crystal structure analysis. 
\vspace{.8mm}
\section*{Competing interests}
The authors declare no competing financial or non-financial interests.
\vspace{1mm}

\section*{references}


%

\widetext
\begin{center}
\newpage
\hspace{0pt}
\textbf{\large Supplementary Material}
\hspace{0pt}
\end{center}
\setcounter{section}{0}
\setcounter{equation}{0}
\setcounter{figure}{0}
\setcounter{table}{0}
\makeatletter
\renewcommand\thesection{S\arabic{section}}
\renewcommand{\theequation}{S\arabic{equation}}
\renewcommand{\thetable}{S\arabic{table}}
\renewcommand\thefigure{S\arabic{figure}}
\renewcommand{\theHtable}{S\thetable}
\renewcommand{\theHfigure}{S\thefigure}
\section{Energy dispersive X-ray spectroscopy}
\begin{figure*}[htbp]
\begin{center}
\includegraphics[width=.8\linewidth]{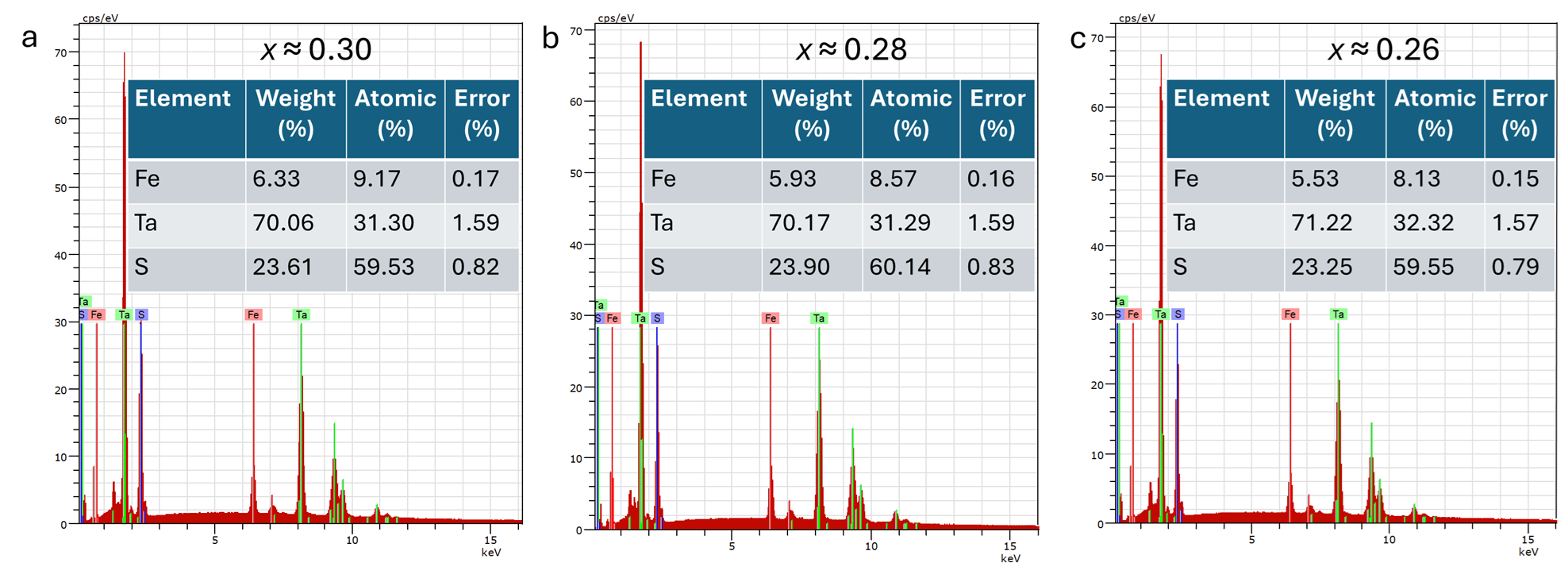}
    \caption{\small \textbf{ Energy dispersive X-ray spectroscopy.} a-c, Energy dispersive X-ray spectroscopy and the atomic percentages of the elements for Fe$_x$TaS$_2$ ($x \approx$ 0.30, 0.28, 0.26).}
    \label{EDS}
    \end{center}
\end{figure*}

Figure \ref{EDS} presents the energy dispersive X-ray spectroscopy (EDS) along with the corresponding atomic percentages of the constituent elements for all samples. The EDS confirms the presence of Fe in all samples, with atomic percentages closely matching the nominal compositions. 

\begin{figure*}[htbp]
\begin{center}
\includegraphics[width=0.8\linewidth]{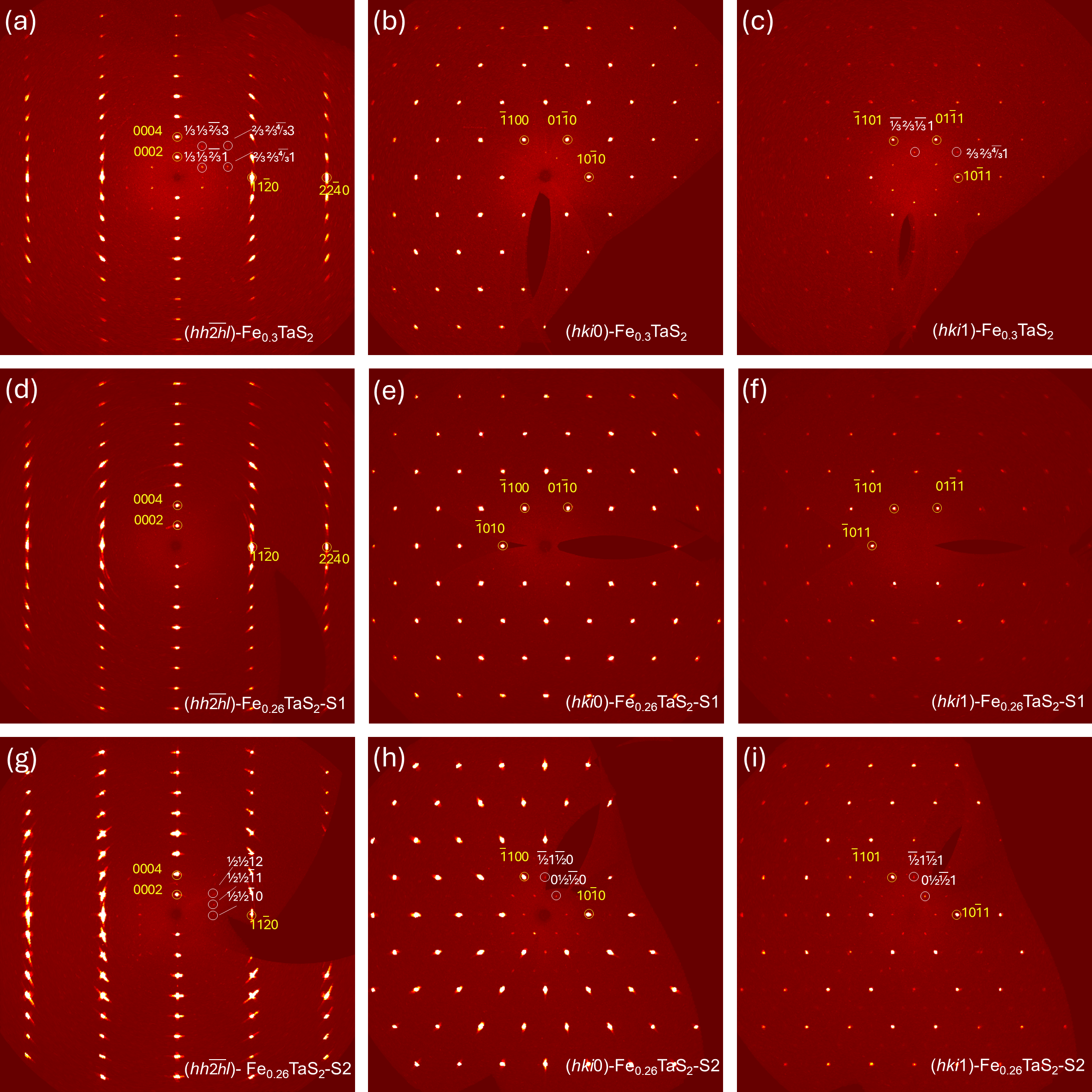}
    \caption{\small \textbf{Single-crystal X-ray diffraction.} a-i, Precession images of the $hh\overline{2h}l$, $hki0$, and $hki1$ planes for an $x \approx 0.30$ sample (a-c) and two distinct $x \approx0.26$ samples (d-f) and (g-i). Note that the Miller indices differ from those used in the main text, as a unified hexagonal unit cell ($a = b = 3.32\ \text{\AA}$, $c = 12.2\ \text{\AA}$, $\alpha = \beta = 90^\circ$, $\gamma = 120^\circ$) was used to generate all precession images, allowing direct comparison between samples regardless of their actual unit cells. Strong integer reflections (yellow circles) correspond to the underlying 2H-$\mathrm{TaS_2}$ lattice. In (a–c), additional weak reflections at one-third and two-thirds positions (white circles) indicate the $\sqrt{3} \times \sqrt{3}$ superlattice of Fe$_{1/3}$TaS$_2$. In contrast, no additional reflections are observed in (d–f), while weak half-integer reflections in (g–i) (white circles) correspond to the $2 \times 2$ superlattice of Fe$_{1/4}$TaS$_2$.}
    \label{Precession Images}
    \end{center}
\end{figure*}

\section{Single Crystal X-ray Diffraction }
Crystal structures of Fe$_x$TaS$_2$ ($x \approx 0.30, 0.26$) are studied by single-crystal X-ray diffraction (SCXRD). For each compound, samples with a typical size of $30\times 30\times 10\ \mathrm{\mu}\mathrm{m}$ are separated from bulk crystals and scanned on a Bruker D8 X-ray diffractometer equipped with APEX2 area detector and Mo K$\alpha$ source ($\lambda = 0.71\ \text{\AA}$). Data integration, unit cell refinement, numerical absorption corrections, and precession image generation are performed using the APEX3 software suite. Specifically, data integrations are carried out with the SAINT program, while absorption corrections are applied using the SADABS. Precession images are generated within APEX3 to visualize reciprocal lattice planes. A custom Python script was developed to remove bad pixels from the frames prior to this step. For direct comparison between samples, all precession images are generated with the hexagonal unit cell $a = b = 3.32\ \text{\AA}$, $c = 12.2\ \text{\AA}$, $\alpha = \beta = 90^{\circ}$, $\gamma = 120^{\circ}$. The actual unit cells are solved by direct methods using the SHELXS \cite{sheldrick2015crystal1} program and subsequently refined by SHELXL refinement package \cite{sheldrick2015crystal1}. The results are listed in Table \ref{tab:lattice-params}.

\begin{table}[htbp]
    \centering
    \caption{Crystal data and refinement parameters for the Fe$_x$TaS$_2$ samples.}
    \label{tab:lattice-params}
    \scriptsize
    \begin{tabular}{lccc}
        \hline
        \textbf{Identification code} & \textbf{Fe$_{0.30}$TaS$_2\ \ \ $} & \textbf{Fe$_{0.26}$TaS$_2$-$S_1$} & \textbf{Fe$_{0.26}$TaS$_2$-$S_2$} \\
        \hline
        Empirical formula           & Fe$_{0.33}$TaS$_2$   & Fe$_{0.30}$TaS$_2$       & Fe$_{0.26}$TaS$_2$\\
        Formula weight (g/mol)      & 264.03               & 261.82                    & 260.71                  \\
        Temperature (K)             & 293                  & 293                       & 293                  \\
        Crystal system              & hexagonal            & hexagonal                 & hexagonal               \\
        Space group                 & P6$_3$22             & P6$_3$/mmc                & P6$_3$/mmc              \\
        a (\AA)                     & 5.7400(3)            & 3.3162(4)                  & 6.6242(6)               \\
        b (\AA)                     & 5.7400(3)            & 3.3162(4)                  & 6.6242(6)               \\
        c (\AA)                     & 12.2401(13)          & 12.226(2)                  & 12.1934(16)              \\
        $\alpha$ (°)                & 90                   & 90                         & 90                      \\
        $\beta$ (°)                 & 90                   & 90                         & 90                      \\
        $\gamma$ (°)                & 120                  & 120                        & 120                     \\
        Volume (\AA$^3$)            & 349.25(5)            & 116.44(4)                  & 463.36(11)               \\
        Z                           & 6                    & 2                           & 8                       \\
        Density (g/cm$^3$)          & 7.517                & 7.468                       & 7.474                   \\
        Absorption Coefficient ($\mu$, mm$^{-1}$) & 50.535 & 50.346                      & 50.485                  \\
        F(000)                      & 312                  & 226                         & 898                    \\
        Reflections Collected       & 12426                & 2545                        & 16221                   \\
        Independent Reflections     & 338                 & 121                         & 310                     \\
        R$_\mathrm{int}$            & 0.1136               & 0.1261                      & 0.0503                  \\
        R$_\sigma$                  & 0.0246               & 0.0412                      & 0.0188                  \\
        Goodness-of-fit on F$^2$    & 1.276                & 1.287                       & 1.225                   \\
        Final R1 [I$\ge$2$\sigma$(I)] & 0.0611             & 0.0561                      & 0.0748                  \\
        Final wR2 [I$\ge$2$\sigma$(I)] & 0.1692            & 0.1140                      & 0.2807                  \\
        Largest diff. peak/hole     & 11.20/-3.38           & 2.65/-2.23                   & 8.05/-5.17              \\
        \hline
    \end{tabular}
\end{table}

Figures \ref{Precession Images}a-c are the precession images for a representative $x \approx 0.30$ sample. The strong integer reflections marked with yellow circles are consistent with the underlying 2H-$\mathrm{TaS_2}$ lattice. The weak reflections, marked with white circles, have fractional indices $h=n\pm1/3,k=m\pm1/3$ and are consistent with the $\sqrt{3} \times \sqrt{3}$ superlattice of Fe$_{1/3}$TaS$_2$. Notably, the fractional reflections with even $l$ are difficult to identify. This suppression arises from a small structure factor rather than a systematic absence. In addition, the largest residue Q peak/hole from refinement is abnormally large, which may originate from the coexistence of the ordered Fe-intercalated phase with a small fraction of disordered intercalated phase, as such a mixture is easily observed in samples with larger sizes. 

For the $x \approx 0.26$ compound, more than 10 samples were examined, and two types of crystal were identified. The first type, which accounts for the majority of the samples, does not exhibit fractional reflections. Figs. \ref{Precession Images}d-f show the precession images for a representative sample S1 of this type. Structural refinement reveals residual charge peaks between $\mathrm{TaS_2}$ layers and the structure is therefore refined with disordered Fe-intercalation. The second type, in contrast, displays weak reflections with half-integer indices. This type is very rare, and Figs. \ref{Precession Images}g-i illustrate the precession images for the only sample S2 showing this behavior. The half-integer reflections, marked with white circles, are consistent with the $2 \times 2$ superlattice of Fe$_{1/4}$TaS$_2$. These observations suggest that the bulk crystal primarily forms a disordered Fe-intercalated phase, despite one single sample with the ordered Fe$_{1/4}$TaS$_2$ phase was identified.

\newpage
\section{Selected Area electron diffraction (SAED) }
\begin{figure*}[htbp]
\begin{center}
\includegraphics[width=.8\linewidth]{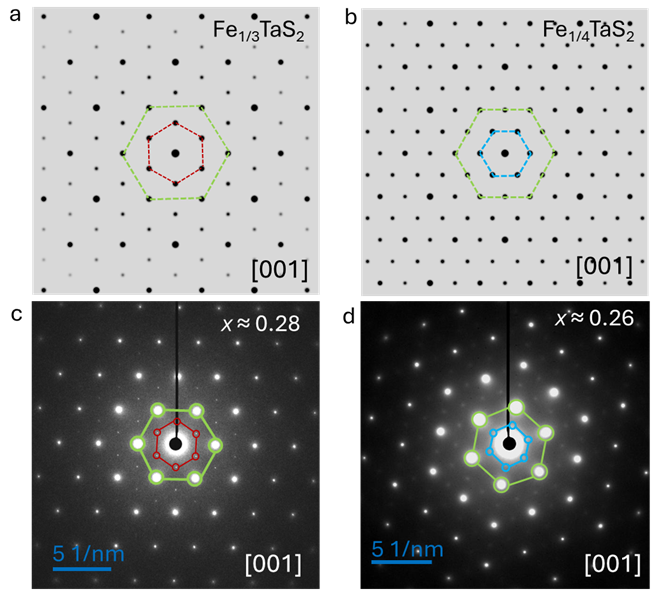}
    \caption{\small \textbf{ Selected area electron diffraction (SAED)}. a,b, Simulated electron diffraction pattern along the [001] for Fe$_{1/3}$TaS$_2$ and Fe$_{1/4}$TaS$_2$, respectively. c,d, Experimental selected area electron diffraction pattern along the [001] for $x \approx 0.28$ and $x \approx 0.26$, respectively.}
    \label{SAED Spplementary}
    \end{center}
\end{figure*}

Figure \ref{SAED Spplementary}a,b shows the simulated electron diffraction pattern along the [001] for Fe$_{1/3}$TaS$_2$ (a) and Fe$_{1/4}$TaS$_2$ (b). The TaS$_2$ unit cell is highlighted by the green dashed line, the $\sqrt{3} \times \sqrt{3}$ superlattice unit cell is marked by the red dashed line, and the $2 \times 2$ superlattice unit cell is indicated with the blue dashed line. The experimental SAED pattern for $x \approx 0.28$ along the [001] direction is shown in Fig. \ref{SAED Spplementary}c. The strong diffraction spots are associated with the TaS$_2$ unit cell, while the weaker diffraction spots (outlined by red circles) correspond to the superlattice spots. The superlattice spots are rotated and form a $\sqrt{3} \times \sqrt{3}$ superlattice, which belongs to the non-centrosymmetric structure of space group P6$_3$22. For $x \approx 0.26$ (Fig. \ref{SAED Spplementary}d), the superlattice spots (outlined by blue circles) formed a $2 \times 2$ superlattice that belongs to the centrosymmetric structure of space group P6$_3$/mmc.

\newpage
\section{Rotation anisotropy second harmonic generation (RA-SHG)}

\begin{figure*}[htbp]
\begin{center}
\includegraphics[width=0.7\linewidth]{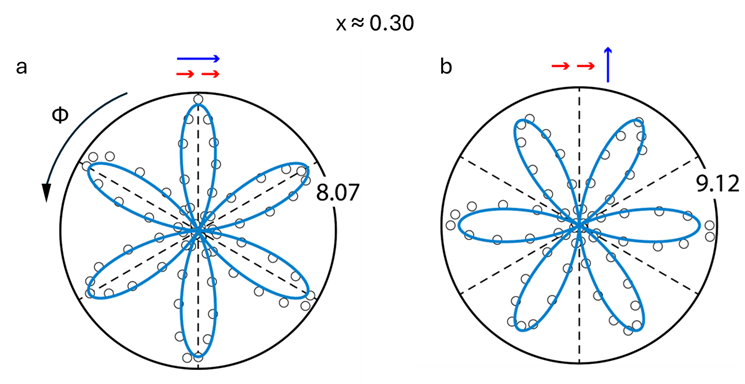}

    \caption{\small \textbf{Polar plots for the intensity of rotation anisotropy second harmonic generation (RA-SHG) for $x \approx 0.30$.} a,b, RA-SHG polar plots taken in parallel (a) and cross channels (b) for $x \approx 0.30$. Arrows representing polarization of SHG (blue) and incident light (red) at $0^o$ (i.e., a-axis). The relative scale of each polar plot is labeled on the edge; 1.0 represents 57,000 photon counts integrated over 5 seconds.}
    \label{SHG}
    \end{center}
\end{figure*}

We performed rotation anisotropy (RA) second harmonic generation (SHG) \cite{jin2020observation1,ahn2024electric1} measurements to examine the status of inversion symmetry in Fe$_{0.30}$TaS$_2$. Figures \ref{SHG}a and \ref{SHG}b show the RA-SHG polar plots of the SHG intensity ($I^{2\omega}$) as a function of the incident polarization angle ($\phi$) in both parallel and cross channels, respectively. We can clearly see the RA-SHG patterns preserve three-fold rotational symmetry ($C_3$) with a strong SHG efficiency. 
This suggests that SHG arises from the leading-order bulk electric-dipole contribution, consistent with the breaking of spatial inversion symmetry.

\section{Magnetization, Hall Resistivity and Longitudinal Resistivity }

\begin{figure*}[h!]
\begin{center}
\includegraphics[width=1\linewidth]{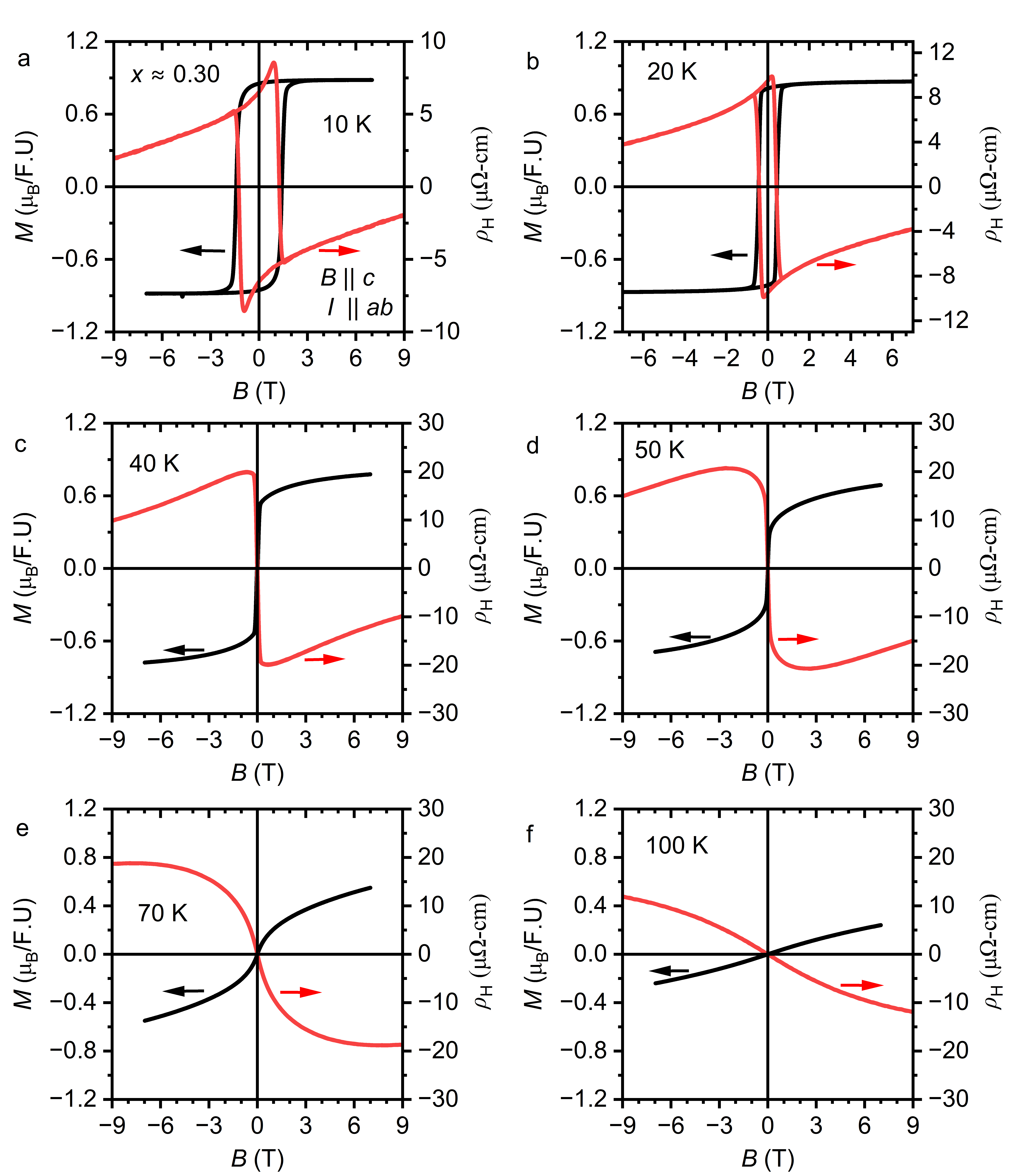}
    \caption{\small \textbf{ Magnetization and Hall resistivity for $x \approx 0.30$.} a-f Magnetization (black curve, left axis) and Hall resistivity (red curve, right axis) plotted as a function of magnetic field for $B||c$ and $I||ab$ at selected temperatures for $x \approx 0.30$.}
    \label{Hall Fe0.3 Supplementary}
    \end{center}
\end{figure*}

\begin{figure*}[h!]
\begin{center}
\includegraphics[width=1\linewidth]{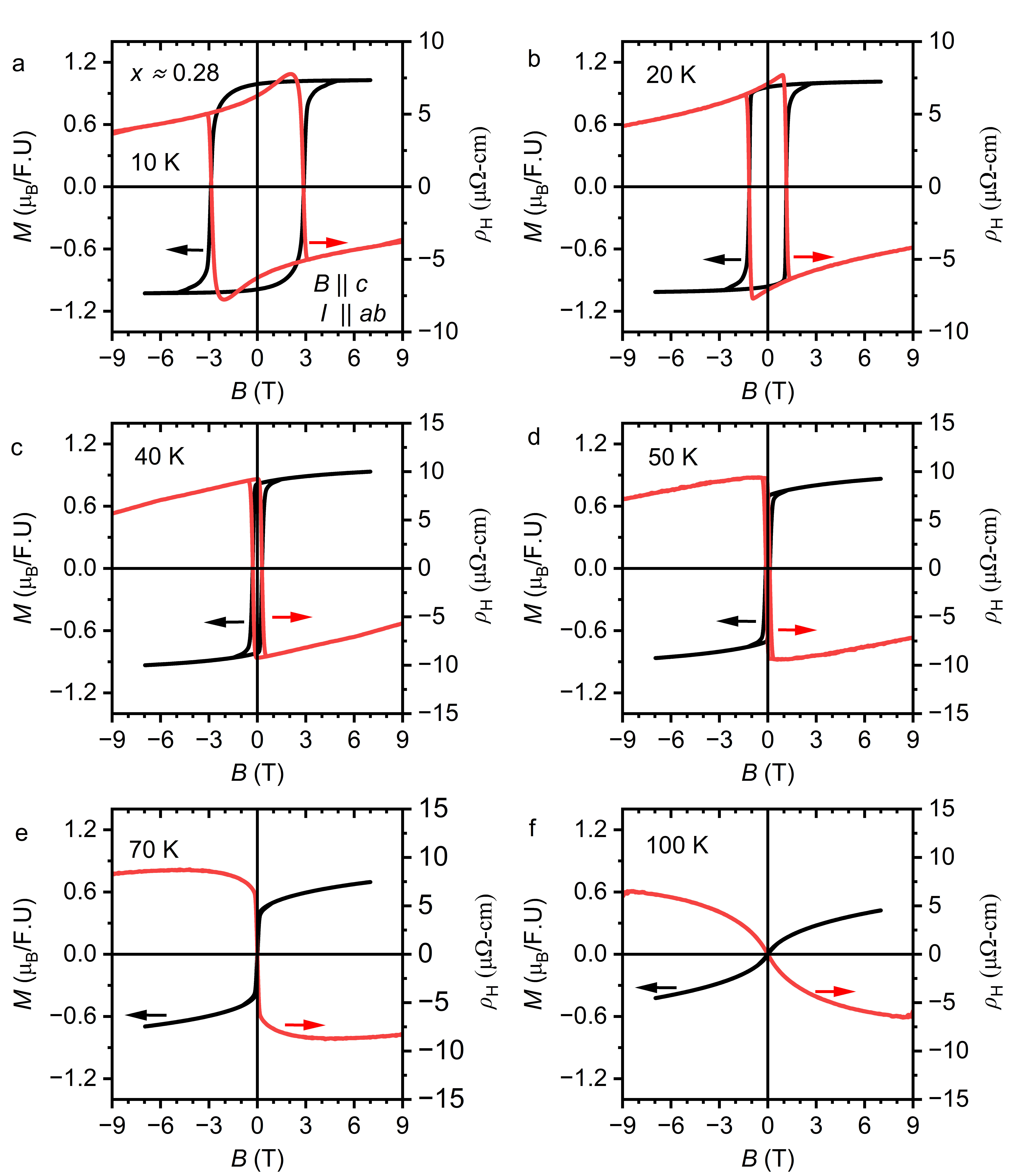}
    \caption{\small \textbf{ Magnetization and Hall resistivity for $x \approx 0.28$.} a-f Magnetization (black curve, left axis) and Hall resistivity (red curve, right axis) plotted as a function of magnetic field for $B||c$ and $I||ab$ at selected temperatures for $x \approx 0.28$.}
    \label{Hall Fe0.28 Supplementary}
    \end{center}
\end{figure*}

\begin{figure*}[htbp]
\begin{center}
\includegraphics[width=1\linewidth]{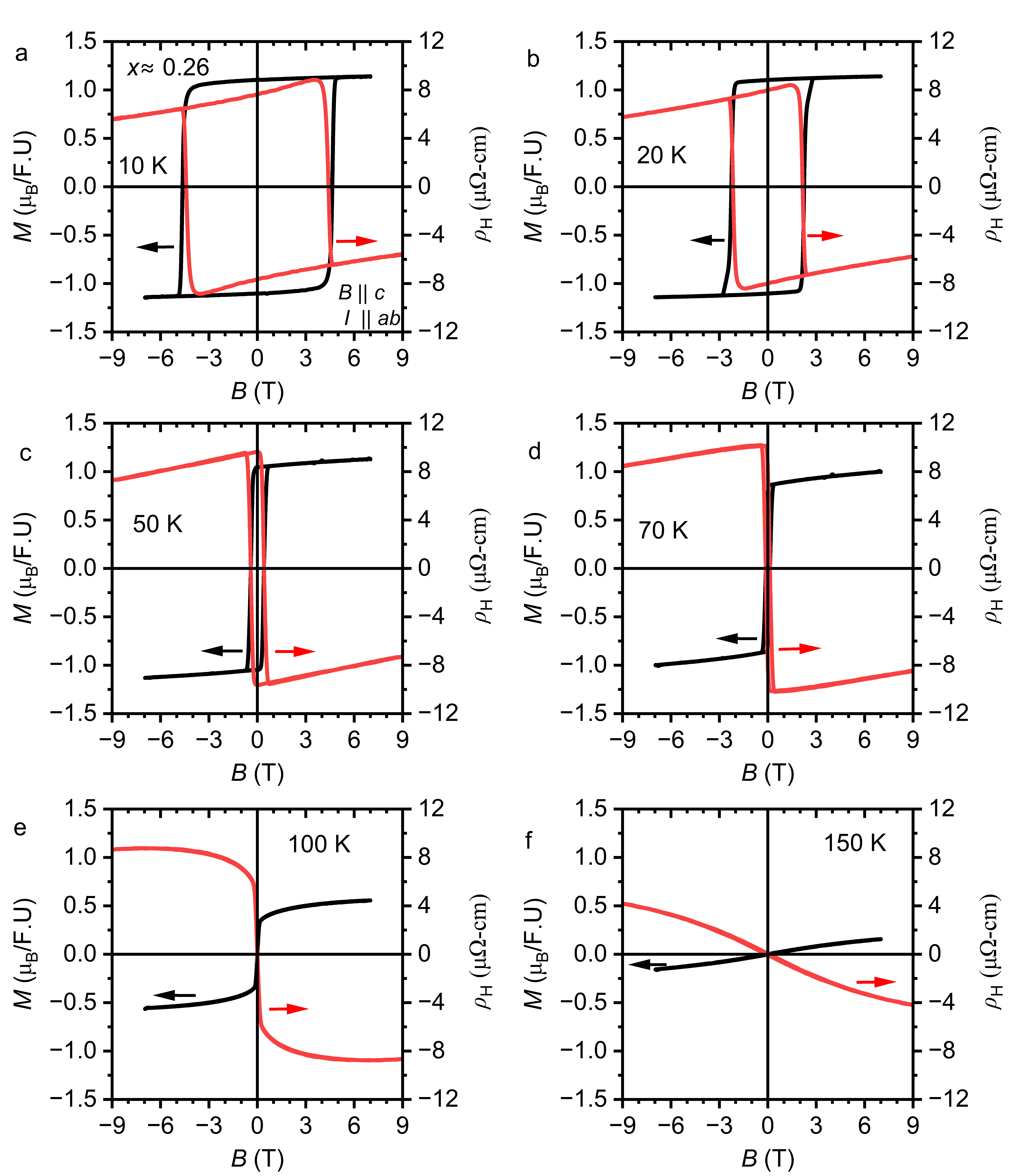}
    \caption{\small \textbf{ Magnetization and Hall resistivity for x $\approx$ 0.26.} a-f Magnetization (black curve, left axis) and Hall resistivity (red curve, right axis) plotted as a function of magnetic field for $B||c$ and $I||ab$ at selected temperatures for x $\approx$ 0.26.}
    \label{Hall Fe0.26 Supplementary}
    \end{center}
\end{figure*}

Figures \ref{Hall Fe0.3 Supplementary}-\ref{Hall Fe0.26 Supplementary} show the magnetization [$M(B)$] and Hall resistivity [$\rho_{H}(B)$] at high temperatures. For $x\approx$ 0.30, the dip-like feature vanishes at and above 30 K (Fig. \ref{Hall Fe0.3 Supplementary}b). Beyond this temperature, the Hall resistivity no longer shows any anomalies and closely follows the trend of the magnetization curve (Figs. \ref{Hall Fe0.3 Supplementary}c-f). For $x \approx$ 0.28, the dip-like feature disappears at and above 30 K (Fig. \ref{Hall Fe0.28 Supplementary}c-f).

\begin{figure*}[htbp]
\begin{center}
\includegraphics[width=.7\linewidth]{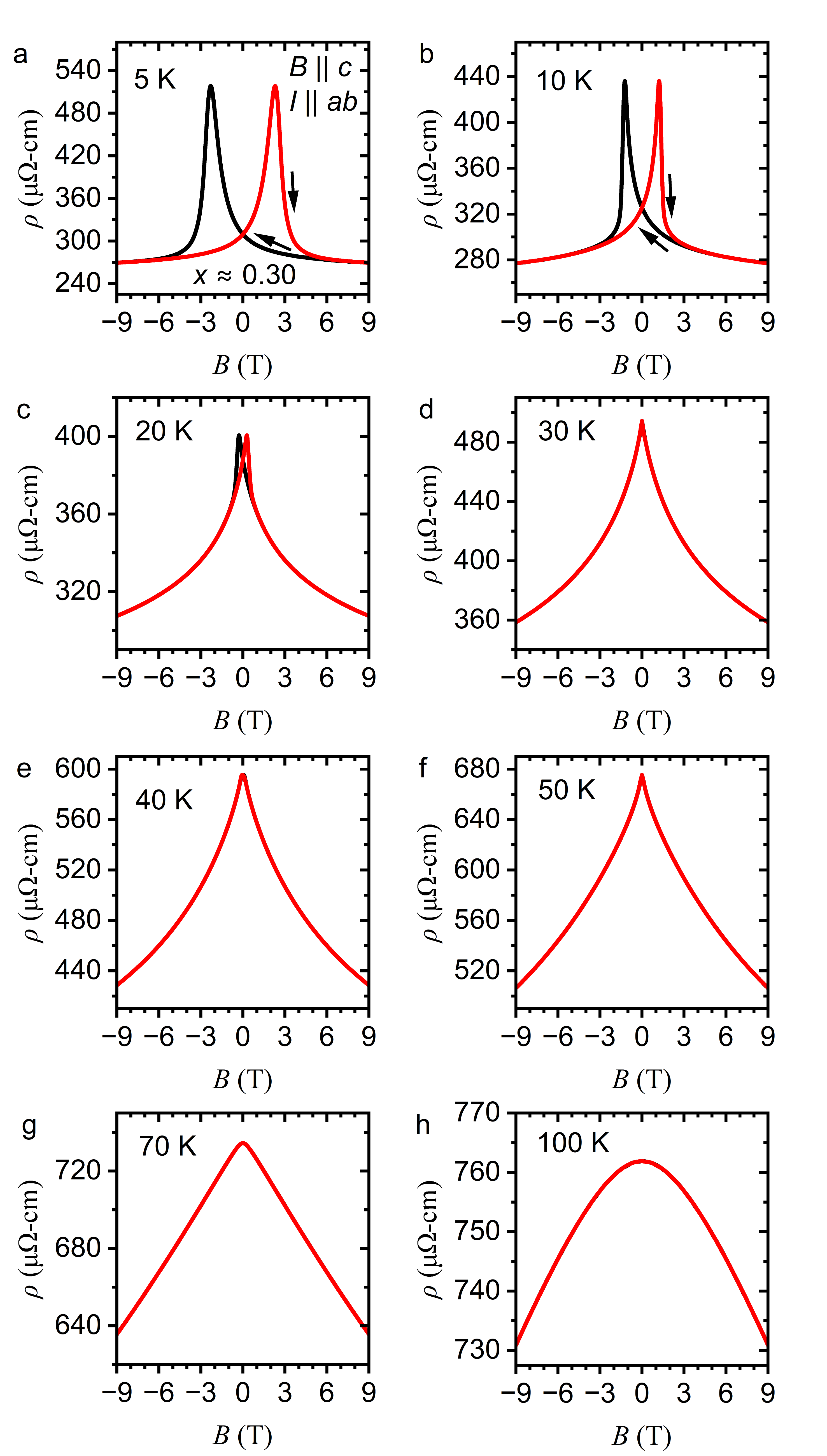}
    \caption{\small \textbf{ Longitudinal resistivity for $x \approx$ 0.30.} a-h Longitudinal resistivity as a function of magnetic field for $B||c$ and $I||ab$ at selected temperatures for $x \approx$ 0.30.}
    \label{rho-H Fe0.30}
    \end{center}
\end{figure*}

\begin{figure*}[htbp]
\begin{center}
\includegraphics[width=.7\linewidth]{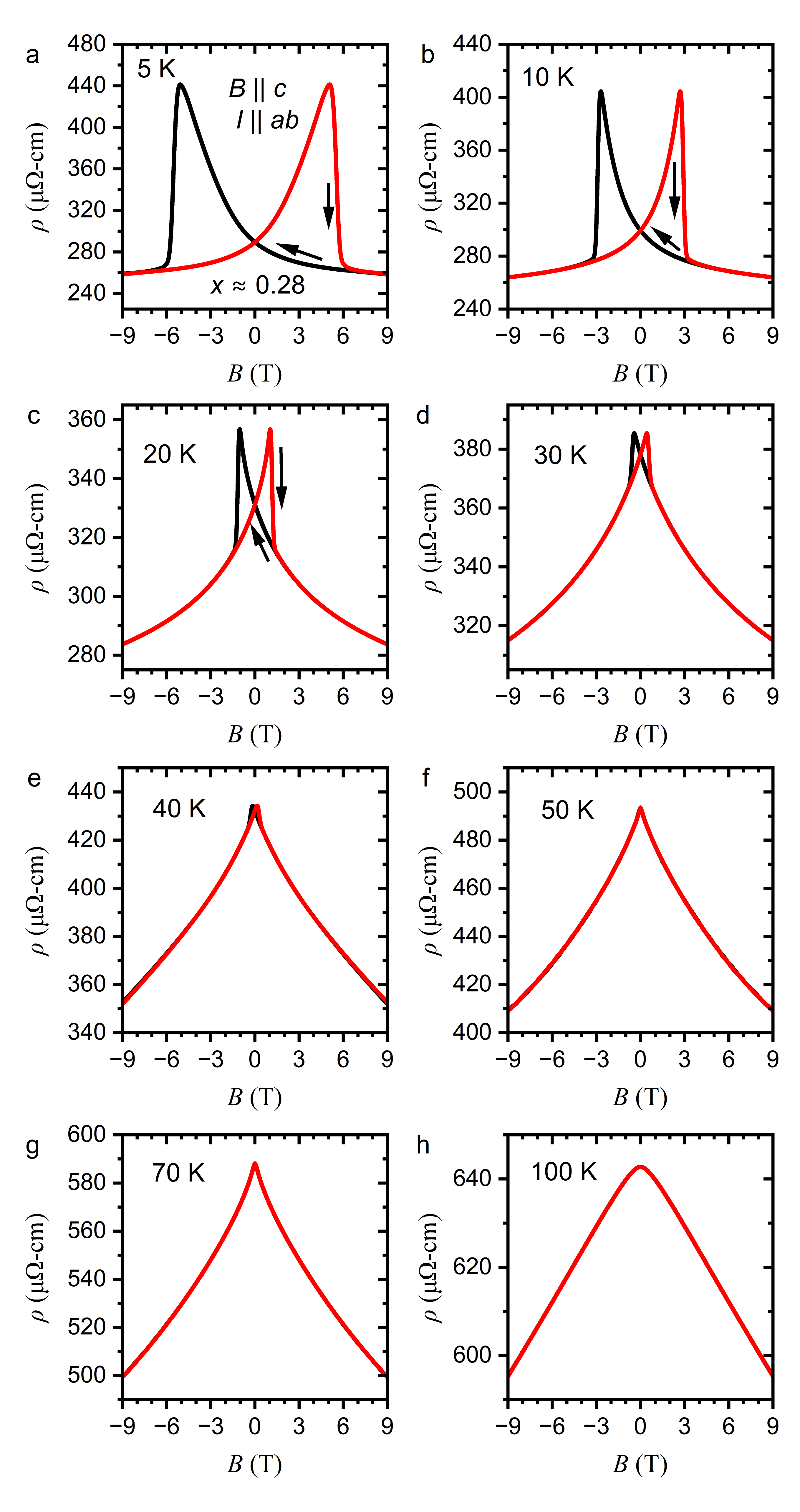}
    \caption{\small \textbf{ Longitudinal resistivity for $x \approx$ 0.28.} a-h Longitudinal resistivity as a function of magnetic field for $B||c$ and $I||ab$ at selected temperatures for $x \approx$ 0.28.}
    \label{rho-H Fe0.28}
    \end{center}
\end{figure*}

\begin{figure*}[htbp]
\begin{center}
\includegraphics[width=.7\linewidth]{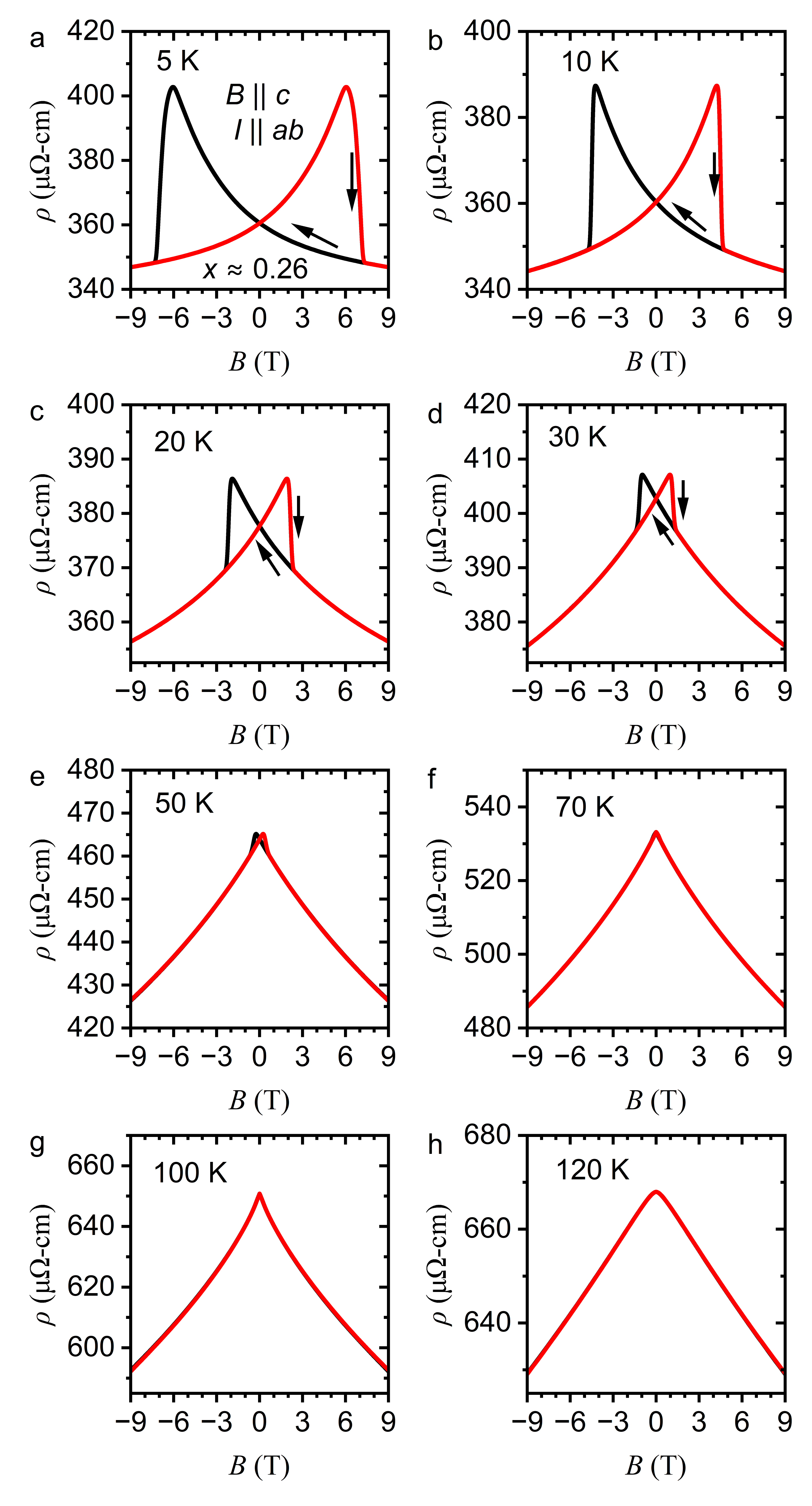}
    \caption{\small \textbf{ Longitudinal resistivity for $x \approx$ 0.26.} a-h Longitudinal resistivity as a function of magnetic field for $B||c$ and $I||ab$ at selected temperatures for $x \approx$ 0.26.}
    \label{rho-H Fe0.26}
    \end{center}
\end{figure*}

\begin{figure*}[htbp]
\begin{center}
\includegraphics[width=1\linewidth]{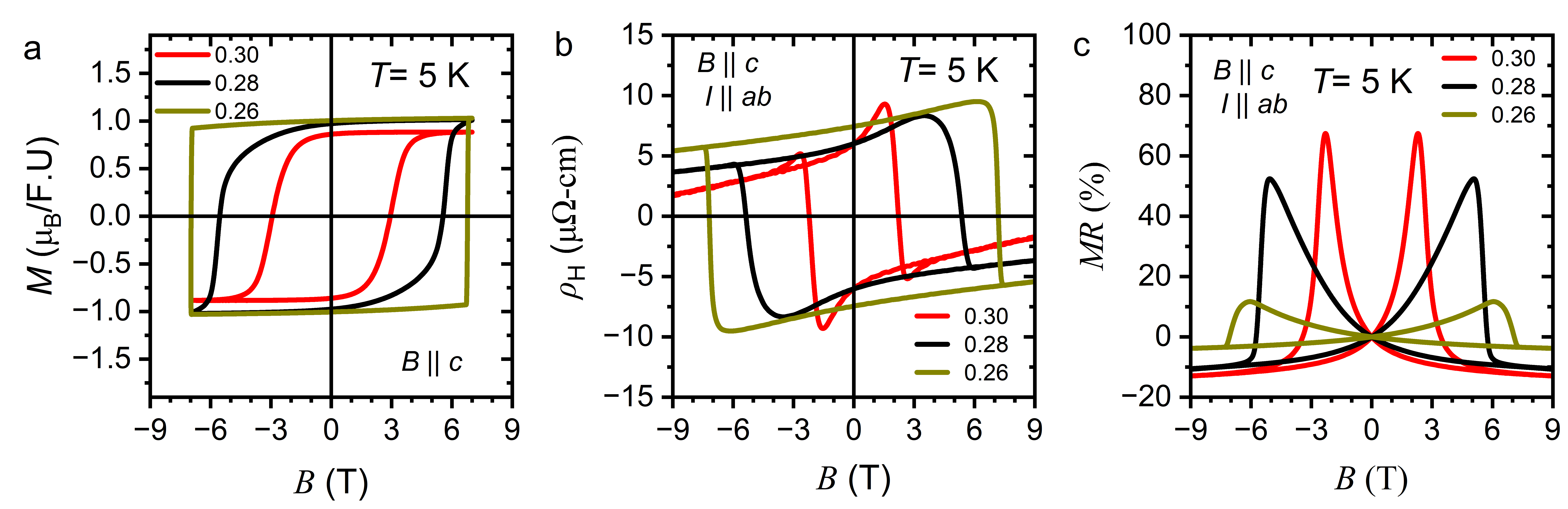}
    \caption{\small \textbf{ Comparison of Magnetization, Hall resistivity, and magnetoresistance.} a-c,  Magnetization (a), Hall resistivity (b), and magnetoresistance (c) vs. magnetic field for Fe$_x$TaS$_2$ (x $\approx$ 0.30, 0.28, 0.26) at 5 K.}
    \label{Comparison}
    \end{center}
\end{figure*}

 In $x \approx$ 0.26, the Hall resistivity closely follows the magnetization curve across all temperatures, without exhibiting any additional features, as shown in Fig. \ref{Hall Fe0.26 Supplementary}. 

The field-dependent longitudinal resistivity [$\rho(B)$] for all compositions is shown in Figs. \ref{rho-H Fe0.30}-\ref{rho-H Fe0.26}. The $\rho(B)$ initially increases as the magnetic field is swept from +9 T toward the negative field direction, followed by a sharp drop at the coercive field. The decrease in $\rho(B)$ observed in the high-field region is typically characteristic of ferromagnetic systems in the saturated magnetic state. Unlike the Hall resistivity, longitudinal resistivity [$\rho(B)$] exhibits identical behavior across all compositions. 

A comparison of magnetization, Hall resistivity, and magnetoresistance at 5 K between all three compositions ($x \approx 0.30, 0.28, 0.26$) is shown in Fig. \ref{Comparison}. The magnetic moment and coercive field increase with decreasing Fe intercalation (Fig. \ref{Comparison}a). At the saturated state, the Hall resistivity is higher for compositions with lower Fe intercalation, consistent with the magnetization behavior (Fig.\ref{Comparison} b). In the case of magnetoresistance ($MR=\frac{\rho(B)-\rho(0)}{\rho(0)}\times 100\%$ ), $x \approx 0.30$ exhibits the highest magnetoresistance.

\section{Magnetic force microscopy (MFM)}
Figure \ref{15 K Negative field sweep} shows the MFM images at higher field at 15 K during the field sweep from +9 T to -9 T. Above -1.1 T only uniform magnetic contrast is observed. 
The field evolution of magnetic domains for the $x \approx 0.30$ sample at 15 K during the field sweep from –9 T to +9 T is shown in Fig. \ref{15 K Negative field sweep}. As expected, uniform magnetic contrast is observed below zero field. At –0.75 T, pronounced stripe domains emerge, coinciding with the field range where a topological Hall feature appears in the Hall resistivity. At higher fields, the stripe domains vanish and uniform magnetic contrast reappears.

\begin{figure*}[h!]
	\centering
	\includegraphics[angle=0,width=14cm,clip]{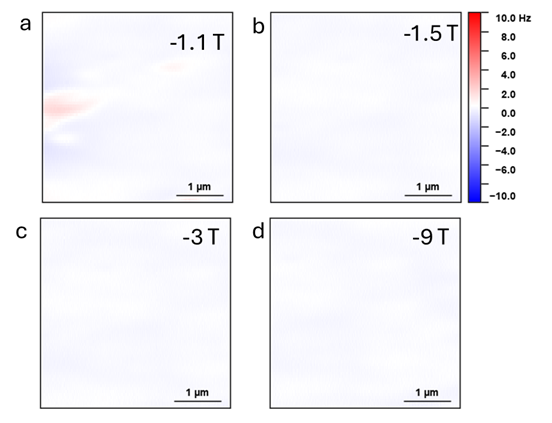}
	\caption{ \textbf{MFM images for $x \approx 0.30$ for high field at 15 K.} a-f, Magnetic domains at high field during field sweep from -9 T to + 9T.
		\label{15 K positive field sweep}}
\end{figure*}

\begin{figure*}[h!]
	\centering
	\includegraphics[angle=0,width=10cm,clip]{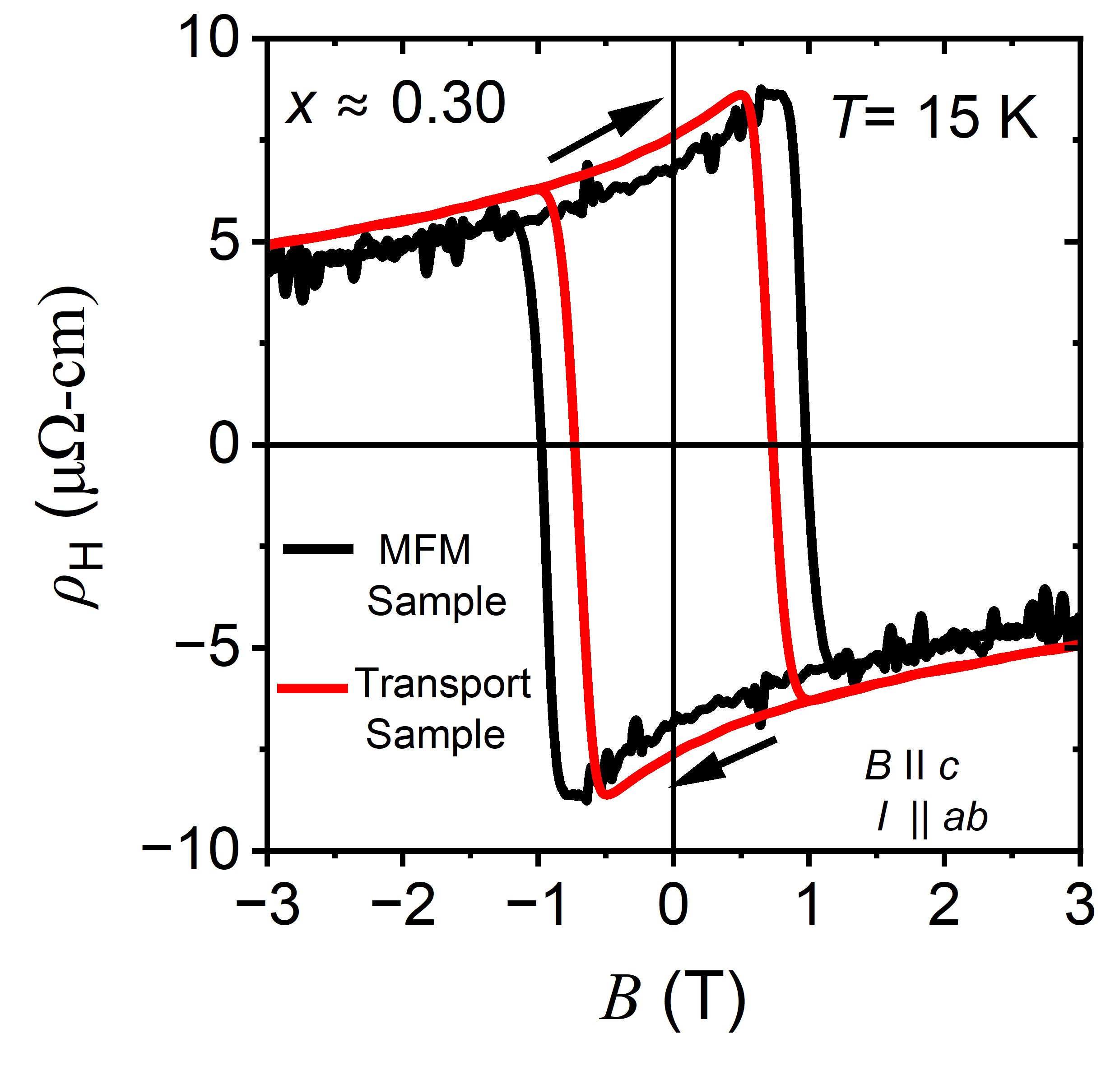}
	\caption{ \textbf{In-situ and direct Hall resistivity at 15 K for $x\approx0.30$.} a, Hall resistivity of MFM sample (black curve) and normal transport sample (red curve) at 15 K. Black arrows denote the field sweep directions.
		\label{In-situ}}
\end{figure*}

\begin{figure*}[h!]
	\centering
	\includegraphics[angle=0,width=16cm,clip]{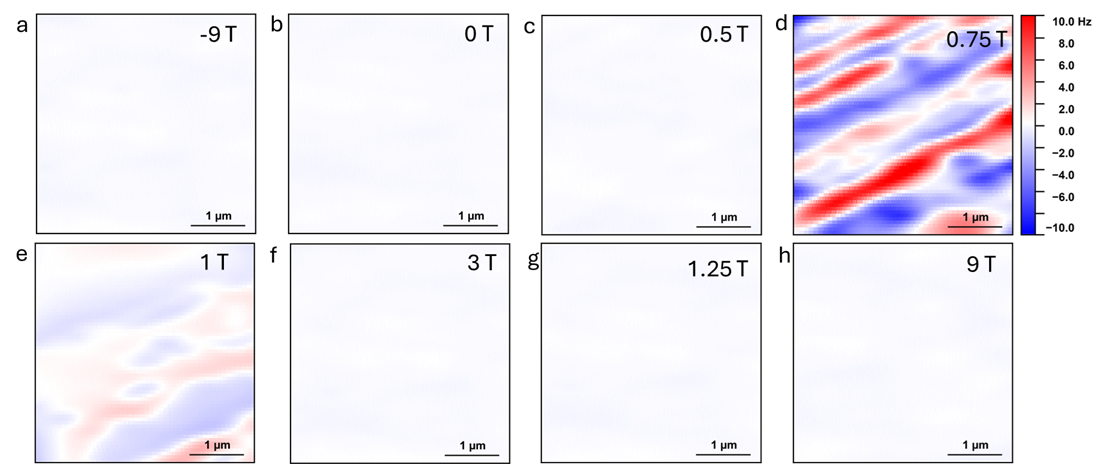}
	\caption{ \textbf{MFM images for $x \approx 0.30$ at 15 K during negative to positive field sweep.} a-f, Field evolution of magnetic domains during field sweep from -9 T to + 9T.
		\label{15 K Negative field sweep}}
\end{figure*}

\begin{figure*}[htbp]
	\centering
	\includegraphics[angle=0,width=16cm,clip]{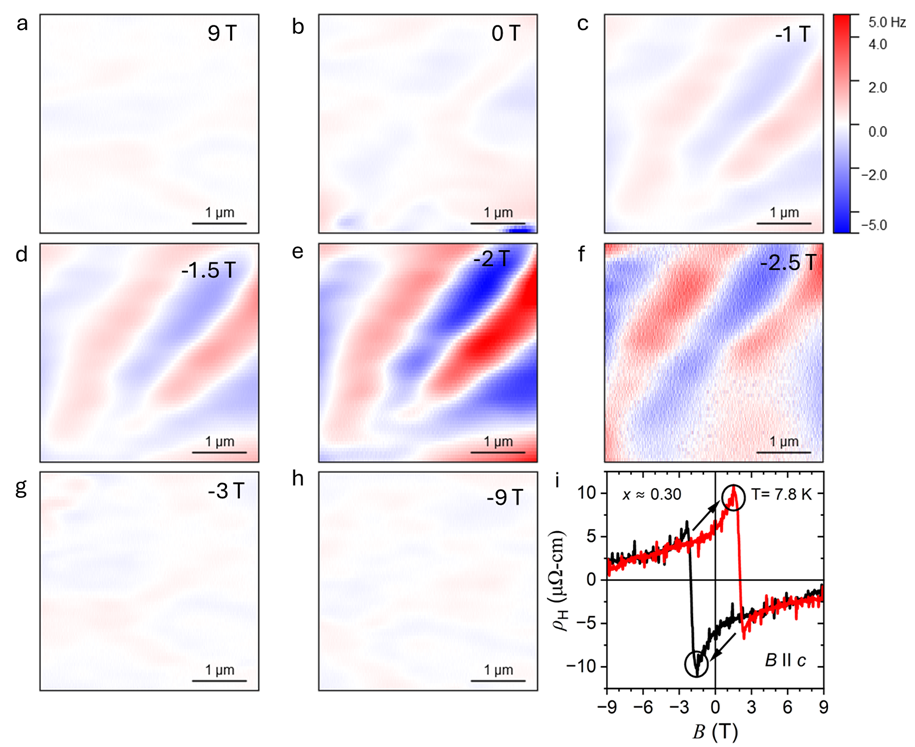}
	\caption{ \textbf{MFM images and in-situ Hall resistivity at 7.8 K for $x \approx 0.30$ during positive to negative field sweep.} a-h, Field evolution of magnetic domains during +9 T to -9 T field sweep.  i,  In-situ Hall resistivity measured simultaneously. Back arrows indicate the field sweep directions. The THE features are outlined by the black circles in the Hall resistivity.
		\label{MFM Fe 0.30 7.8 K}}
\end{figure*}

\begin{figure*}[htbp]
	\centering
	\includegraphics[angle=0,width=14cm,clip]{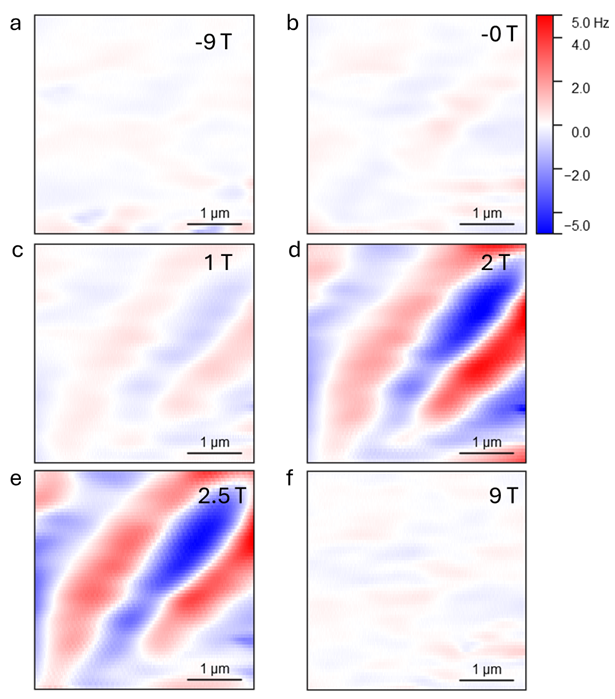}
	\caption{ \textbf{MFM images $x \approx 0.30$ at 7.8 K  during negative to positive field sweep.} a-h, Magnetic domains during field sweep from -9 T to +9 T.
		\label{7.8 K Negative field sweep}}
\end{figure*}

Figures \ref{MFM Fe 0.30 7.8 K}–\ref{7.8 K Negative field sweep} show the MFM images obtained for $x \approx 0.30$ at 7.8 K during positive-to-negative and negative-to-positive field sweeps, respectively. During the field sweep from +9 T to –9 T, an almost uniform magnetic contrast is observed below zero field (Fig. \ref{MFM Fe 0.30 7.8 K}a), while faint magnetic contrast emerges at zero field (Fig. \ref{MFM Fe 0.30 7.8 K}b). At - 1 T, large regularly spaced strip domains develop (Fig. \ref{MFM Fe 0.30 7.8 K}c ). Further decreasing the field, the magnetic contrast increases and the stripe domains persist up to -2.5 T (Figs. \ref{MFM Fe 0.30 7.8 K}d-f). At higher negative fields, the stripe domains disappear (Figs. \ref{MFM Fe 0.30 7.8 K}g,h). The in-situ Hall resistivity measured simultaneously with the MFM imaging is shown in Fig. \ref{MFM Fe 0.30 7.8 K}(i), where topological contributions are evident (outlined by black circles). The field regime of the stripe domains coincides with the topological Hall response, directly linking the two phenomena. A similar behavior is observed during the reverse field sweep from –9 T to +9 T [Fig. \ref{7.8 K Negative field sweep}], where stripe domains reappear just above zero field in the same field range where topological Hall features are detected.

\section{Reproducibility of Hall resistivity}
\begin{figure}[H]
\begin{center}
\includegraphics[width=.7\linewidth]{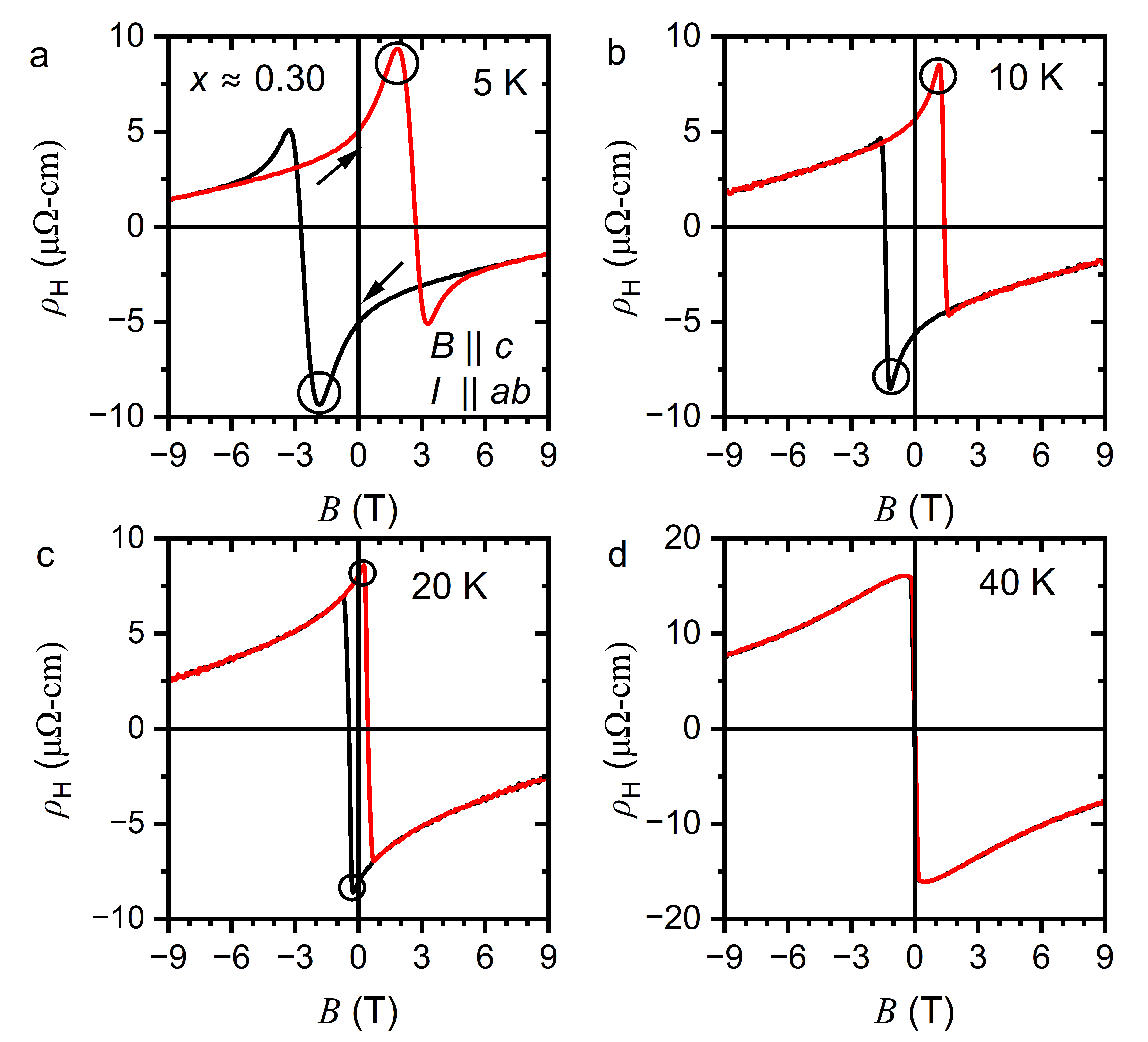}
    \caption{\small \textbf{ Hall resistivity on a second set of single-crystal for $x \approx 0.30$.} a-d Hall resistivity plotted as a function of magnetic field at selected temperatures for $x \approx 0.30$, measured in a second batch of single-crystal. The dip-like features are outlined by black circles, respectively.}
    \label{Fe0.3 Sample2}
    \end{center}
\end{figure}

\begin{figure}[H]
\begin{center}
\includegraphics[width=.8\linewidth]{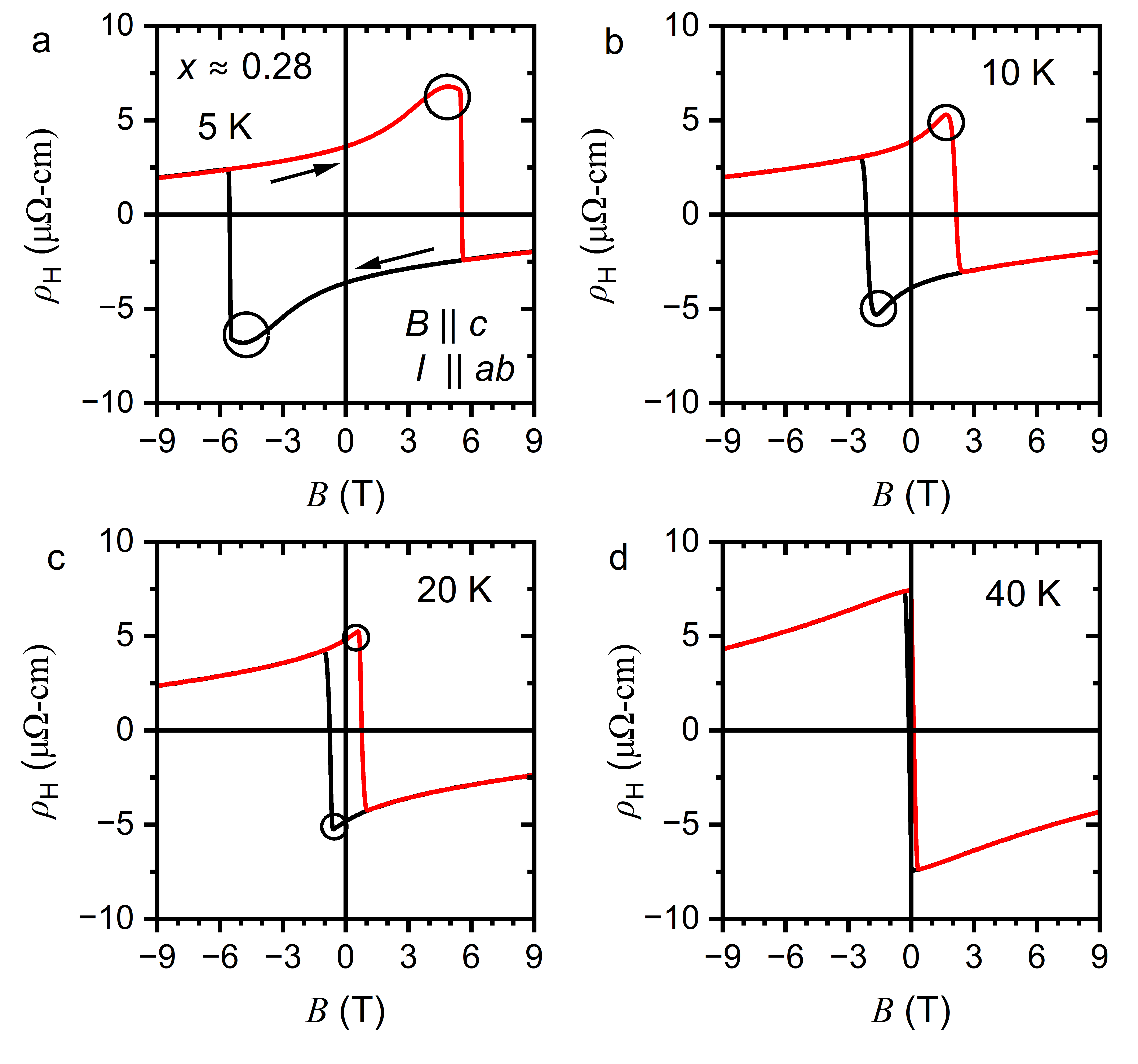}
    \caption{\small \textbf{ Hall resistivity on a second set of single-crystal for $x \approx 0.28$.} a-d Hall resistivity plotted as a function of magnetic field at selected temperatures for $x \approx 0.28$, measured in a second batch of single-crystal. The dip-like feature is shown by the black circles.}
    \label{Fe0.28 Sample 2}
    \end{center}
\end{figure}

\begin{figure}[H]
\begin{center}
\includegraphics[width=.8\linewidth]{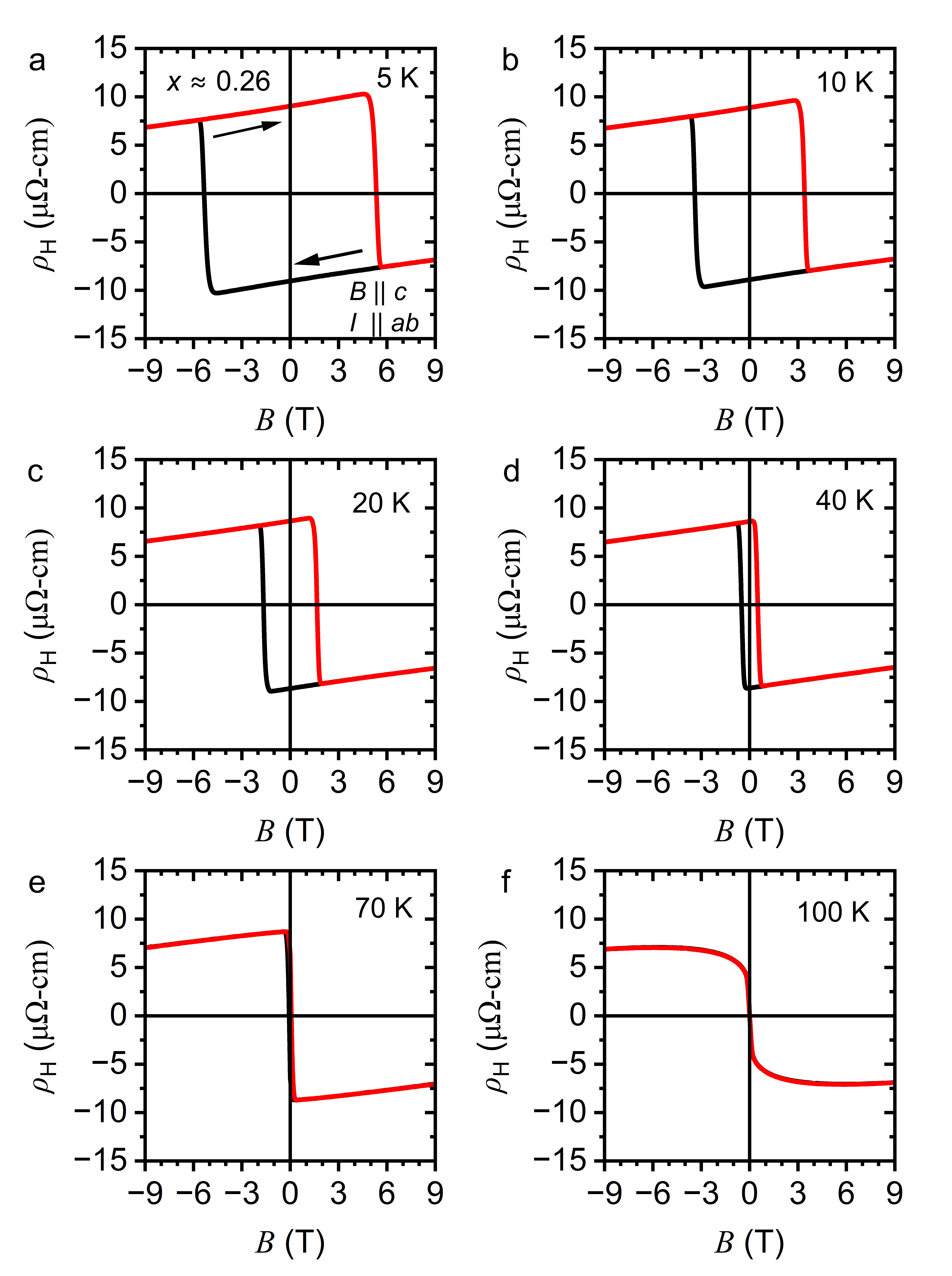}
    \caption{\small \textbf{ Hall resistivity on a second set of single-crystal for $x \approx 0.26$.} a-d Hall resistivity plotted as a function of magnetic field at selected temperatures for $x \approx 0.26$, measured in a second batch of single-crystal.}
    \label{Fe0.26 Sample 2}
    \end{center}
\end{figure}

The Hall resistivity is reproduced in several independently grown batches of single crystals. Figures \ref{Fe0.3 Sample2}-\ref{Fe0.26 Sample 2} show the Hall resistivity for $ x \approx 0.30, 0.28$, and $0.26$, measured in a second batch of single crystals grown separately. The dip-and hump-like features are observed for $ x \approx 0.30$, whereas only a dip-like feature is observed for $ x \approx 0.28$. As expected, $ x \approx 0.26$ does not show any such behavior. The reproducibility of the Hall resistivity across different batches of crystals indicates that the observed phenomena are robust and intrinsic to the material.

\clearpage
\section*{Supplementary References}

\end{document}